\documentclass{emulateapj}
\usepackage{apjfonts}
\usepackage{amssymb}
\usepackage{amsmath}
\usepackage{amsbsy}
\usepackage{epsfig}

\def\tpm{\theta_{\rm PM}}
\def\tsis{\theta_{\rm SIS}}
\def\tx{\theta_x}
\def\ty{\theta_y}
\def\tc{\theta_{\rm C}}
\def\tb{\theta_{\rm b}}

\begin{document}

\title{Gravitational Lensing Signatures of Supermassive Black Holes in Future Radio Surveys}

\author{Judd D. Bowman\altaffilmark{1}, Jacqueline N. Hewitt\altaffilmark{1},
James R. Kiger\altaffilmark{1}}

\email{jdbowman@mit.edu; jhewitt@space.mit.edu}

\altaffiltext{1}{MIT Center for Space Research, 77 Massachusetts
Ave., Cambridge, MA 02139}


\begin{abstract}
Observational measurements of the relationship between supermassive
black holes (SMBHs) and the properties of their host galaxies are an
important method for probing theoretical hierarchical growth models.
Gravitational lensing is a unique mechanism for acquiring this
information in systems at cosmologically significant redshifts. We
review the calculations required to include SMBHs in two standard
galactic lens models, a cored isothermal sphere and a broken power
law.  The presence of the SMBH produces two primary effects depending
on the lens configuration, either blocking the ``core'' image that is
usually predicted to form from a softened lens model, or adding an
extra, highly demagnified, image to the predictions of the unaltered
lens model. The magnitudes of these effects are very sensitive to
galaxy core sizes and SMBH masses. Therefore, observations of these
lenses would probe the properties of the inner regions of galaxies,
including their SMBHs. Lensing cross-sections and optical depth
calculations indicate that in order to fully observe these
characteristic signatures, flux ratios of order $10^6$ or more
between the brightest and faintest images of the lens must be
detectable, and thus the next generation of radio telescope
technology offers the first opportunity for a serious observational
campaign.  Core images, however, are already detectable and with
additional observations their statistics may be used to guide future
SMBH searches.
\end{abstract}


\section{INTRODUCTION}
\label{sec_introduction}

The central regions of galaxies contain a great deal of information
regarding their formation and evolution. The mass-density profiles of
these inner regions have been recognized for sometime as important
measures of cold dark matter models \citep{1997ApJ...490..493N,
2001ApJ...549L..25K}, and more recently, the properties of central
supermassive black holes (SMBHs) have emerged as important probes
into the history of hierarchical growth models.

SMBHs are inferred compact objects with masses between $10^6$ and
$10^9$ solar masses.  Several empirical correlations have been
identified between SMBH masses and the properties of their host
galaxies, including correlations with the mass or luminosity of the
galactic bulge \citep{1998AJ....115.2285M, 2000AAS...196.2122K,
2001ApJ...553..677L}, with the bulge velocity dispersion
\citep{2000ApJ...539L..13G, 2001cksa.conf..335M}, and also with the
light concentration of the bulge \citep{2001ApJ...563L..11G}. Recent
analysis suggests that SMBH mass may additionally be correlated with
the circular velocity of the host galaxy, even beyond the optical
radius \citep{2002ApJ...578...90F, 2003MNRAS.341L..44B}, and thus
there may be a link between the SMBH mass the total mass of the
galaxy. Because of this apparently tight link between SMBHs and their
host galaxies, including dark matter halos, it is likely that the
formation of both were closely related. Most contemporary
hierarchical growth models do in fact reflect this property
\citep{1998A&A...331L...1S, 2000MNRAS.311..576K, 2001ApJ...554L.151B,
2002ApJ...569...83M, 2002ApJ...581..886W, 2003ApJ...582..559V,
2003ApJ...593...56D}.

The detection of SMBHs is difficult and traditionally has been
achieved using stellar kinematics in the very central regions of
galaxies.  These techniques have been feasible only for relatively
nearby galaxies ($D \lesssim 100$ Mpc, $z \lesssim 0.02$)
\citep{2002ApJ...574..740T} in which the cores can be resolved. SMBHs
are presumed to be the inactive relics of active galactic nuclei, and
thus, have evolved over a long redshift history ($z \gtrsim 6$ to
present).  Motivated, in part, by the desire to observe this
evolution, additional techniques for deducing SMBH mass in more
distance galaxies have been developed. The most prominent of these is
the use of gravitationally broadened emission lines in active
galactic nuclei to infer SMBH mass in these objects
\citep{2001MNRAS.327..199M}.

Gravitational lensing offers a compelling mechanism to explore the
relationship between the properties of SMBHs and their host galaxies
at intermediate redshifts, extending to $z \simeq 1$ or greater.
Examination of strong galaxy-galaxy (or galaxy-quasar) gravitational
lenses under proper circumstances may yield the mass of the SMBH,
information on the inner mass-density profile of the host galaxy, and
the total mass of the galaxy simultaneously, greatly simplifying
existing observational hurdles.

Relatively minimal consideration has been given to the gravitational
lensing properties of SMBHs in galaxies, however.
\citet{2001ApJ...551..223A} have considered the SMBH in the Milky
Way, discussing the influence of stellar neighbors on its lensing
properties, as well as the effects of its gravitational shear on the
microlensing of individual stars.  For galaxy-galaxy lensing, only
\citet{2001MNRAS.323..301M} and \citet{2003ApJ...587L..55C} have made
general analyses.  The former demonstrate calculations for adding a
SMBH to a standard galactic lensing potential and the latter shows
how SMBHs can affect image separations in lenses.
\citet{2003ApJ...582...17K} considers a sample population of 73 mock
lenses and calculates lensing statistics including SMBHs effects. He
is interested in observing core images, which contain information
about the inner mass-density profiles of galaxies, and finds that
SMBHs should reduce the probability of observing the faintest of
these images, but otherwise not generally affect core images.  Using
observational data, \citet{2001ApJ...549L..33R} show that explicitly
including SMBHs in models weakens constraints on mass profiles in the
inner regions of galaxies and \citet{2004Natur.427..613W} have
achieved an upper bound of $2 \times 10^8$ solar masses for any SMBH
in the lens system, PMN~J1632-0033. \citet{1992ApJ...397L...1W}
discussed using VLBI observations to detect SMBHs in galaxy halos,
and \citet{2001PhRvL..86..584W} and \citet{2001MNRAS.320L..40A}
placed limits on the abundance of supermassive compact objects in the
universe.

In this paper we expand upon these initial efforts and provide a
general context for interpreting SMBH effects in lensing studies.  In
particular, we will elaborate on the discussions of
\citet{2003ApJ...582...17K} and \citet{2001MNRAS.323..301M} regarding
the impact of SMBHs on observing core images and also perform
calculations to determine the likelihood of directly observing SMBH
lensing signatures. We begin in Section \ref{sec_analytic} by
presenting two common galactic lens models and performing the
calculations to describe the addition of a SMBH to each. From these
modified models we establish criteria which a lens-source system must
meet in order for SMBH effects to be detectable. In Section
\ref{sec_numerical} we use these criteria to compute cross-sections
for the production of observable lensing events. Using the observed
local relation between SMBH mass and bulge velocity dispersion, these
cross-sections are convolved with the velocity dispersion
distribution of galaxies to establish estimated optical depths to
lensing.  We conclude, in Sections \ref{sec_empirical} and
\ref{sec_conclusion}, by summarizing the lensing signatures of SMBHs
and discussing their effects on other lensing studies, such as the
search for core images, and also how SMBH signatures themselves might
be observed with existing and proposed future facilities.

\section{ANALYTICAL TREATMENT OF LENS MODELS}
\label{sec_analytic}

\subsection{Lensing by a Single Galaxy with a SMBH}

In order to determine the statistics of lensing by a supermassive
black hole in a galaxy core, we must first address the mechanics of
the lensing. In this section we review the properties of combined
galaxy and black hole lensing, emphasizing those aspects that are
important for the determination of cross-sections for lensing
statistics.  The SMBH is modeled as a point mass and we consider two
models for the potential of a lensing galaxy, a cored isothermal
sphere and a broken power law. It has been recognized for some time
that the observable surface brightnesses in the central regions of
galaxies do not actually approach zero slope as implied by the
canonical cored isothermal potential. This has been the motivation
for various broken power law parameterizations for surface
brightness, such as the ``Nuker Law'' \citep{1995AJ....110.2622L,
1996AJ....111.1889B, 1997AJ....114.1771F, 1997MNRAS.287..525Z}. While
surface brightness has indeed been shown to be well represented by a
broken power law, the actual mass-density distribution remains
largely unknown and hence we will treat both models in parallel for
this paper.

\subsubsection{Point Mass and Cored Isothermal Sphere (PM+CIS)}

We begin by considering a point mass potential added to a cored
isothermal sphere potential. A similar treatment has already been
performed by \citet{2001MNRAS.323..301M} starting from a cored
isothermal sphere mass-density.  The distinction is subtle and the
resulting behavior is quite similar. The advantage to the approach
followed here is that the characteristic equations are
straightforward.

Assuming the lensing occurs in the weak field limit, the standard
one-dimensional lens equation is \citep[][Section
5.1]{book_gravitational_lenses}
    \begin{equation}
    \label{eqn_lens}
    \beta = \theta - \nabla \Psi,
    \end{equation}
where $\theta$ is the position of the image, $\beta$ is the position
of the source as measured from the center of the lens, $\Psi$ is the
two-dimensional projected potential of the lens, and $\nabla \Psi$
represents the deflection angle.  For an isolated point mass,
    \begin{equation}
    \Psi = \tpm^2 \ln \theta,
    \end{equation}
and for an isolated galaxy with a cored isothermal lensing potential,
    \begin{equation}
    \Psi = \tsis \sqrt{\theta^2 + \tc^2}
    \end{equation}
\citep[][Table 2]{astro-ph/9606001}.  The core radius of the galaxy
potential is denoted by $\tc$ and the Einstein radii of the black
hole and galaxy are denoted by $\tpm$ and $\tsis$, respectively:
    \begin{eqnarray}
    &&\theta_{\rm PM} = \sqrt{\frac{4GM_{\rm BH}}{c^2} \frac{D_{\rm LS}}{D_{\rm OS}D_{\rm OL}}} \label{eqn_tpm} \\
    &&\theta_{\rm SIS} = \frac{4 \pi \sigma^2}{c^2}\frac{D_{\rm LS}}{D_{\rm OS}}
    \label{thetasis}.
    \end{eqnarray}
The velocity dispersion of the galaxy is $\sigma$, the mass of the
black hole is $M_{\rm BH}$, and the distances from observer to lens,
lens to source, and observer to source are $D_{\rm OL}$, $D_{\rm
LS}$, and $D_{\rm OS}$, respectively.

For the weak field limit, lensing potentials add linearly, thus a
reasonable projected potential for lensing by a galaxy with an SMBH
is obtained by simply combining these two potentials to give
    \begin{equation}
    \Psi = \tsis \sqrt{\theta^2 + \tc^2} + \tpm^2 \ln \theta, \label{eqn_softlenspot}
    \end{equation}
and results in the lens equation,
    \begin{equation}
    \beta = \theta - \tsis\frac{\theta}{\sqrt{\theta^2+\tc^2}} - \frac{\tpm^2}{\theta}. \label{softlenseqn}
    \end{equation}
For circularly symmetric lens potentials the magnification at any
image position is given by \citep[][Section
2.3]{book_gravitational_lenses}
    \begin{equation}
    \frac{1}{\mu}=\frac{\beta}{\theta}\frac{\partial \beta}{\partial
    \theta},
    \end{equation}
where $\mu$ is the magnification. The critical curves in the image
plane can be found by setting this equation equal to zero
($\mu\rightarrow\infty$), which yields two conditions:
    \begin{equation}
    \frac{\beta}{\theta} = 1 - \frac{\tsis}{\sqrt{\theta^2 + \tc^2}} - \frac{\tpm^2}{\theta^2} = 0
    \end{equation}
and
    \begin{equation}
    \frac{\partial \beta}{\partial \theta} = 1 - \frac{\tsis}{\sqrt{\theta^2 + \tc^2}} + \frac{\tsis \theta^2}{\left ( \tc^2 + \theta^2 \right )^{3/2}} + \frac{\tpm^2}{\theta^2} =
    0.
    \end{equation}
The first condition produces a single caustic point at the origin of
the source plane, while the second condition produces two concentric
caustic circles.  In the image plane, the first condition is
responsible for the outer of two critical curves, and the second
condition for the inner plus a critical point at the origin.  Figure
\ref{fig_pos_mag} illustrates the transformation from the CIS
potential to the PM+CIS potential in terms of the image positions and
magnifications as a function $\beta$.

\subsubsection{Point Mass and Broken Power Law Isothermal Sphere
(PM+POW)}

We now derive a similar set of expressions for lensing by a point
mass potential added to a circularly symmetric broken power law.  In
this case, the surface mass density is a more natural starting point
and is given by \citep{2003ApJ...582...17K}
    \begin{equation}
    \Sigma_{POW} \left ( \theta \right ) = 2^{\left ( \eta - \gamma
    \right ) / \alpha} \Upsilon I_{b} \left (
    \frac{\theta}{\theta_{b}} \right ) ^ {-\gamma} \left [ 1 + \left
    ( \frac{\theta}{\theta_{b}} \right ) ^ \alpha \right ] ^ {\left (
    \gamma - \eta \right ) / \alpha},
    \end{equation}
where $\gamma$ and $\eta$ are inner and outer power law indices,
respectively, $\tb$ is the break radius between the power laws,
$I_{b}$ is the surface brightness at the break radius, $\alpha$ gives
the sharpness of the break, and $\Upsilon$ is the mass-to-light
ratio. This yields \citep{2003ApJ...582...17K} a lensing deflection
angle,
    \begin{eqnarray}
    \nonumber \lefteqn{\nabla \Psi = \frac{2^{1 + \left ( \eta - \gamma \right ) /
    \alpha}}{2 - \gamma} \frac{\Upsilon I_{b}
    \theta_{b}}{\Sigma_{cr}} \left ( \frac{\theta}{\theta_{b}} \right
    ) ^ {1 - \gamma} \times} \\
    & & \qquad \qquad \mbox{} \times { }_{2} F_{1} \left [ \frac{2 - \gamma}{\alpha},
    \frac{\eta - \gamma}{\alpha}; 1 + \frac{2 - \gamma}{\alpha}; -
    \left ( \frac{\theta}{\theta_{b}} \right ) ^ \alpha \right ],
    \end{eqnarray}
where $\Sigma_{cr}$ is the critical surface density for lensing and
${}_{2} F_{1}$ is the hypergeometric function.  This model has
introduced several new parameters that are not present in the cored
isothermal sphere. Later, we shall wish to explore the effects of
these new parameters and compare the outcome back to the cored
isothermal model. To facilitate that effort we now restrict the outer
power index to its isothermal equivalent, $\eta=1$, and recognize
that, under this condition, in the limit $\theta_{b} \rightarrow 0$
the surface mass density must match that for the singular isothermal
sphere. Doing so, we have
    \begin{equation}
    \lim_{\theta_{b} \rightarrow 0} \Sigma_{POW} \left ( \theta:\eta=1 \right ) = \Sigma_{SIS} \left ( \theta \right
    ).
    \end{equation}
Using
    \begin{equation} \Sigma_{SIS} \left ( \theta \right ) =
    \frac{\sigma^2}{2 G D_{L} \theta} = \frac{\tsis c^2 D_{OS}}{8 \pi GD_{LS} D_{L}
    \theta},
    \end{equation}
we find the relation
    \begin{equation}
    2^{\left ( 1 - \gamma \right ) / \alpha} \Upsilon I_{b}
    \theta_{b} = \frac{\tsis c^2 D_{OS}}{8 \pi G D_{LS} D_{L}}
    \end{equation}
and are able to write the deflection angle for the combined point
mass and broken power law in terms of $\tsis$:
    \begin{eqnarray}
    \nonumber \lefteqn{\nabla \Psi = \frac{\tpm^2}{\theta} + \frac{\tsis}{2 - \gamma} \left ( \frac{\theta}{\theta_{b}} \right ) ^ {1 - \gamma} \times } \\
    & & \qquad \mbox{} \times {}_{2} F_{1} \left [ \frac{2 - \gamma}{\alpha}, \frac{1 - \gamma}{\alpha}; 1 +
    \frac{2 - \gamma}{\alpha}; - \left ( \frac{\theta}{\theta_{b}} \right ) ^ \alpha \right ].
    \end{eqnarray}

\subsubsection{Typical Values of $\tpm$, $\tc$, $\tb$, $\alpha$, and $\gamma$}
\label{sec_typical_values}

The empirical correlation in the local universe between the masses of
SMBHs and the velocity dispersions of their host galaxy bulges can be
exploited to determine a typical value of $\tpm/\tsis$.  Observable
lenses are dominated by elliptical lensing galaxies
\citep{1984ApJ...284....1T}.  For elliptical galaxies we can ignore
the distinction between the velocity dispersion of the bulge and the
galaxy as a whole. Therefore, we assign the SMBH its mass according
to \citep{2001cksa.conf..335M}:
    \begin{equation}
    \label{eqn_mbh}
    M_{\rm BH} = 1.48 \pm 0.24 \times 10^8 {\rm M}_\odot \left ( \frac{\sigma}{200~{\rm km/sec}} \right )^{4.7},
    \end{equation}
where $\sigma$ is equal to the (bulge) velocity dispersion and
$M_{\odot}$ is the solar mass.  If we temporarily assume that this
relationship holds for all galaxies, even beyond the local universe,
we can generate characteristic curves for $\tpm/\tsis$ as a function
of $\sigma$, $D_{OL}$, $D_{OS}$, and $D_{LS}$ by substituting the
above result into Equation \ref{eqn_tpm}. Figure \ref{fig_pm_to_sis}
illustrates the behavior of $\tpm/\tsis$ for different lens and
source galaxy distances. For lenses located roughly halfway to the
source, which because it is most efficient for lensing will dominate
\citep[Figure 5]{1984ApJ...284....1T}, there is not much variation in
$\tpm/\tsis$ for a fixed velocity dispersion. Galaxies with different
velocity dispersions will produce different $\tpm/\tsis$ ratios.

There is to date no empirical understanding of any relationship
between galaxy cores and other observables similar to that for SMBHs.
The parameterizations we have chosen use angular measures for the
core and break radii which, when mapped back to physical lengths, are
dependent on the redshifts of the lens and source galaxies,
complicating their interpretations.  Figure \ref{fig_rc_to_tc} gives
the relationship between the angular and linear radii.  (From here on
we will simply refer to $\tc$ and $\tb$ as the core radius regardless
of the model under discussion.)

For the PM+POW model, we have already restricted the outer power law
index, $\eta=1$, to match the isothermal potential, but left $\gamma$
and $\alpha$ unconstrained. Figure \ref{fig_alpha_gamma_sequence}
illustrates the effects of varying these parameters. Drawing on the
reported "Nuker Law" model fits, we can calculate average values for
these parameters.  In Table \ref{tab_nuke_average} we compute the
mean and median values for the 26 "cored" galaxies listed in
\citet[][Table 2]{1997AJ....114.1771F}. 
    \begin{deluxetable}{lcc}
    \tablewidth{3in}
    \tablecaption
    {
        Average values for the ``Nuker Law'' parameters
        \label{tab_nuke_average}
    }
    \tablecomments
    {
        These averages were calculated for the ``cored''
        galaxies listed in \citet[][Table 2]{1997AJ....114.1771F}.
        Those authors use $\beta$ to indicate the outer power law
        index, whereas we have used $\eta$.
    }
    \tablehead{\colhead{Parameter} & \colhead{Mean} & \colhead{Median}}

    \startdata
    $\alpha$\dotfill            & 2.35 & 2.03 \\
    $\gamma$\dotfill            & 0.09 & 0.08 \\
    $\eta$\dotfill              & 1.28 & 1.33 \\
    $\tb$ (arcsec)\dotfill      & 1.71 & 1.29 \\
    \enddata

    \end{deluxetable}

%
Based on these findings, we will take $\gamma=0.1$ and $\alpha=2.0$
for all subsequent calculations.  From Table \ref{tab_nuke_average},
it is also apparent that forcing $\eta=1$, as we did above, was not a
bad approximation to these observational results.

\subsubsection{Shear and Ellipticity}

We constrain our analysis to circularly symmetric lensing potentials
for this paper in order to better illustrate the dependence on
fundamental parameters.  As it is well known, however, gravitational
lenses seem to rarely display such simple symmetry and are often best
modeled by elliptical mass density potentials or by including a
gravitational shear in the local environment.  In Appendix A we
present a justification for neglecting the contribution of these
higher order factors in our exploration of the fundamental behavior
of these systems.

\subsection{Signatures of SMBH Lensing}

The addition of a SMBH to an otherwise typical galactic lensing
potential produces three significant effects: 1) two-image lensing
extends infinitely beyond the Einstein radius of the galaxy,  2) the
maximum number of images increases from three to four, and 3) the
area of the region in the source plane expected to produce more than
two images is reduced compared to a cored or broken power law
galactic potential with no SMBH. In general, for most source
positions, $\beta$, the lensing characteristics of the combined SMBH
and galaxy act as if they were merely superimposed, and only conspire
in the central regions of the source plane to cancel out the
secondary images that either would have produced alone.

\begin{figure*}
\begin{center}
\includegraphics{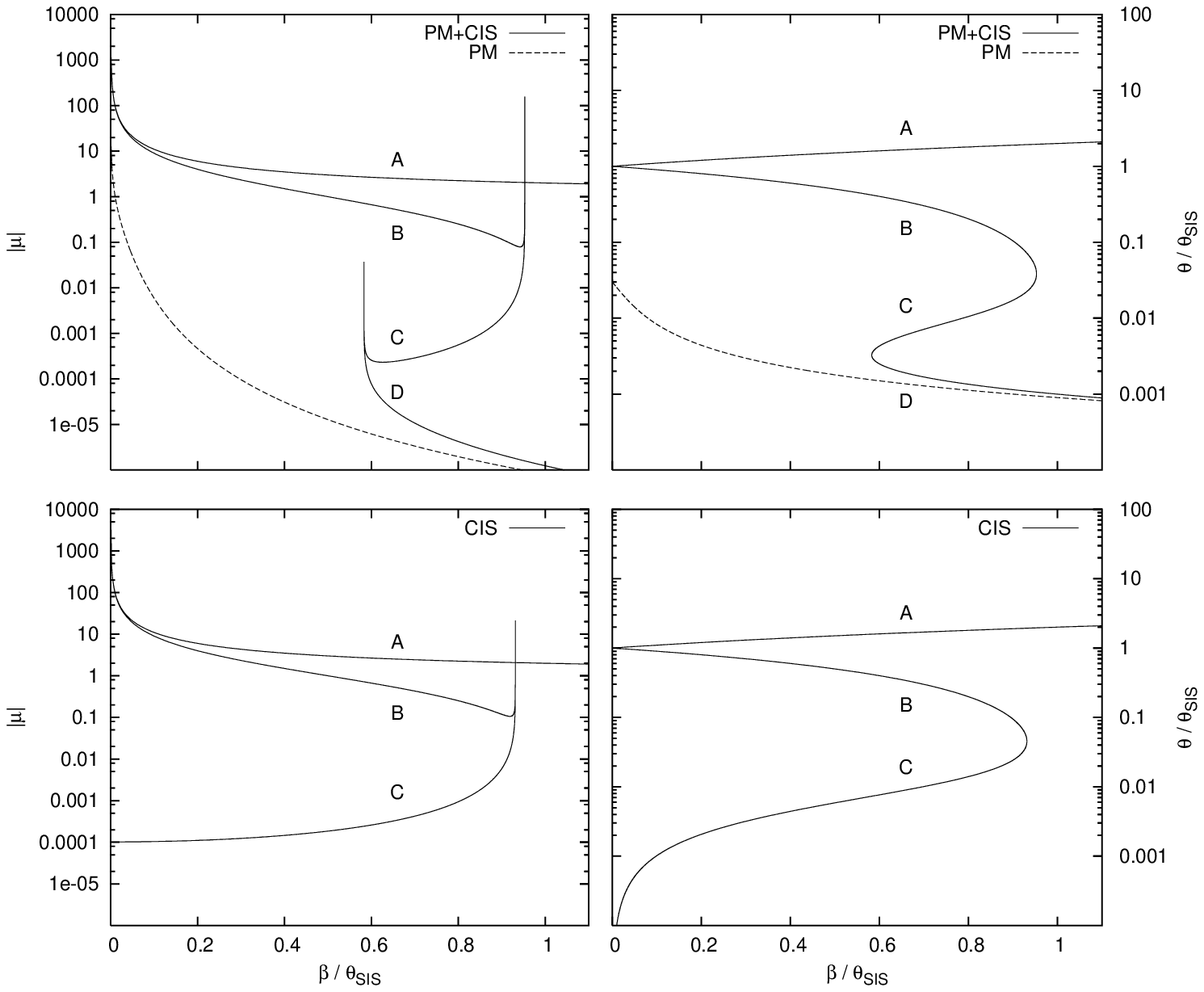}
\end{center}
\caption{ \label{fig_pos_mag} Positions and magnifications of the
images produced by PM+CIS model (top) compared to the cored
isothermal sphere (bottom). Image positions and the source position
are given in units of the singular isothermal sphere Einstein ring
radius, $\tsis$. For this figure we chose $\tc/\tsis=0.01$ and
$\tpm/\tsis=0.03$. The primary image is labelled ``A`` and the
secondary images are labelled ``B,'' ``C,'' and in the case with the
point mass, also ``D.''  The dashed lines indicate the properties of
the secondary image of an isolated point mass. }
\end{figure*}

\begin{figure}
\begin{center}
\includegraphics{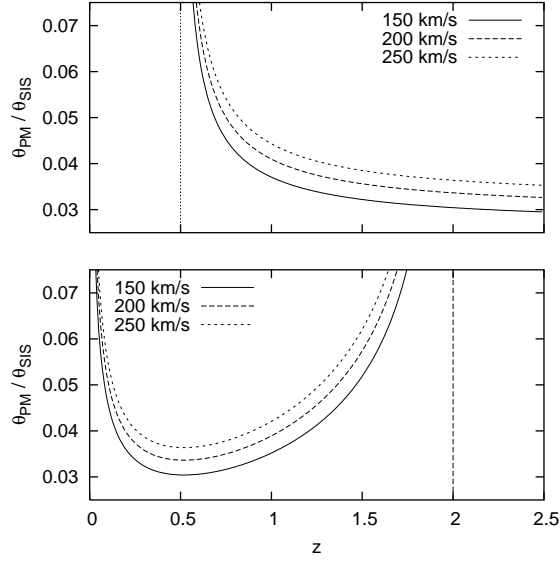}
\end{center}
\caption{\label{fig_pm_to_sis} Einstein radii ratio $\tpm/\tsis$ is
plotted as a function of source redshift for a lens galaxy at
$z_L=0.5$ (top), and as a function of lens redshift for a source
galaxy at $z_S=2.0$ (bottom).  The empirical relationship between
SMBH mass and host galaxy velocity dispersion is used and three
values of velocity dispersion are shown. A standard cosmology of
$\Omega_M=0.3$, $\Omega_\Lambda=0.7$, and $H_0=70$km/s/Mpc is used.}
\end{figure}

\begin{figure}
\begin{center}
\includegraphics{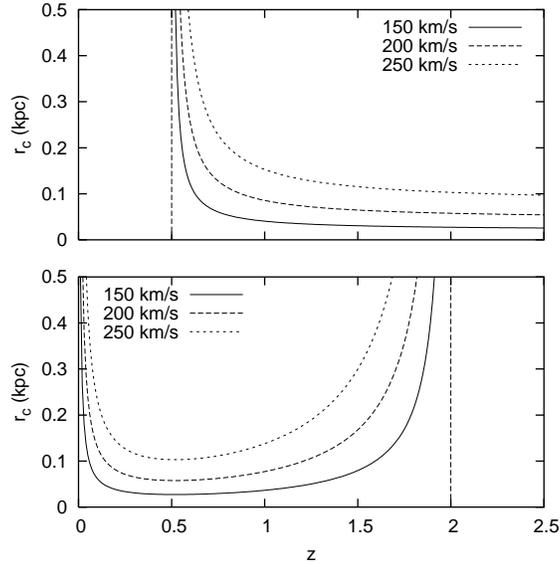}
\end{center}
\caption{\label{fig_rc_to_tc} Physical distance, $r_c$, corresponding
to the angular core radius ratio, $\tc/\tsis=0.1$, is plotted as a
function of source redshift for a lens galaxy at $z_L=0.5$ (top), and
as a function of lens redshift for a source at $z_S=2.0$ (bottom).
Other values of the ratio can easily be deduced because the curves
scale linearly with $\tc/\tsis$. Three values of velocity dispersion
are shown.  A standard cosmology of $\Omega_M=0.3$,
$\Omega_\Lambda=0.7$, and $H_0=70$km/s/Mpc is used.}
\end{figure}

\begin{figure*}
\begin{center}
\includegraphics{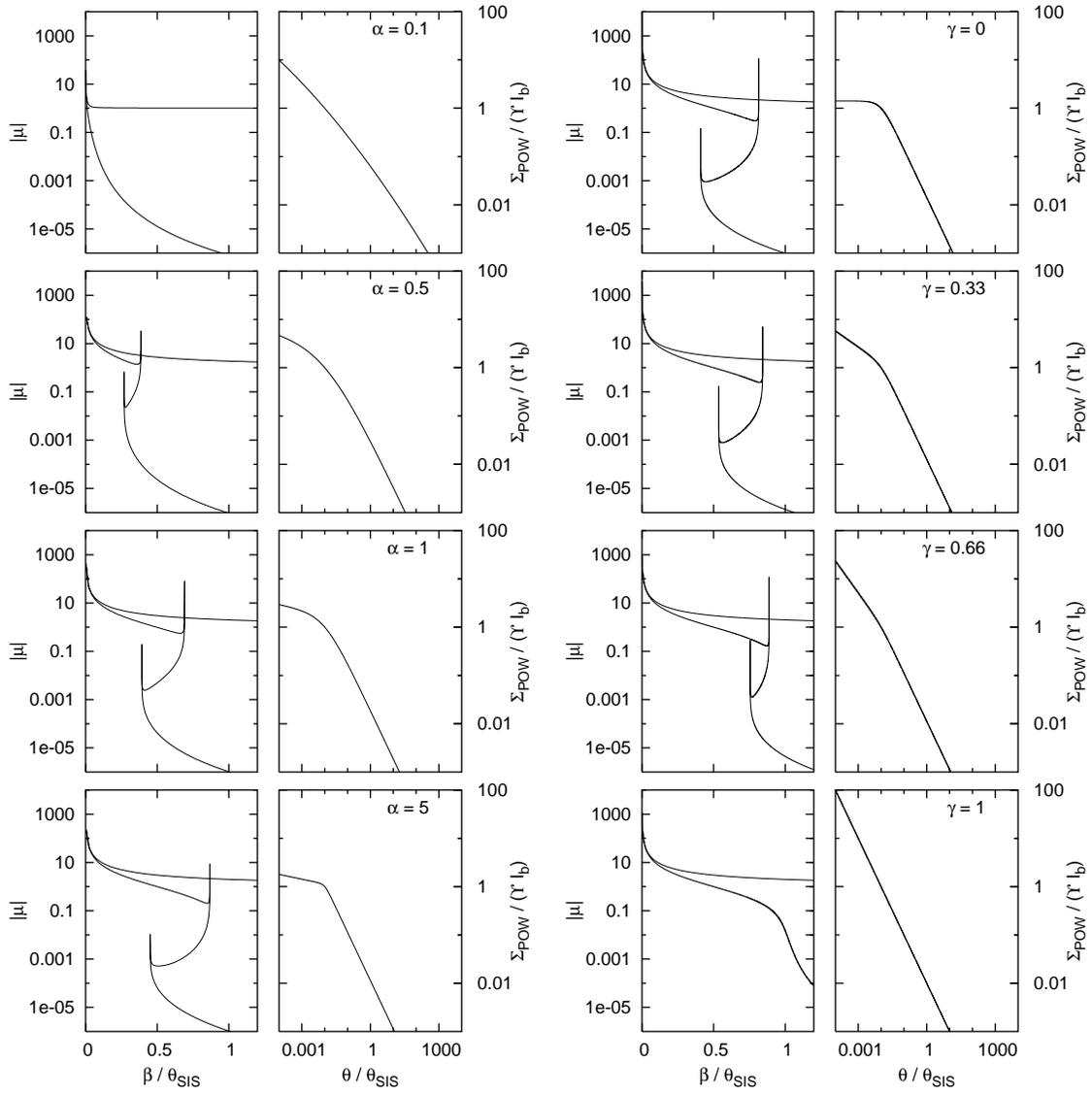}
\end{center}
\caption{\label{fig_alpha_gamma_sequence} Series of plots indicating
the influence of PM+POW parameters on lensing.  The left column is a
series of plots illustrating the progression, from top to bottom, of
the magnification curves for the PM+POW potential as the sharpness of
the break, $\alpha$, is increased while holding constant $\eta=1$,
$\gamma=0.1$, $\tb/\tsis=0.01$, and $\tpm/\tsis=0.03$.  The profile
of the normalized surface brightness, $\Sigma_{POW}/I_b$, is drawn
along side for reference. On the right, the progression illustrates
increasing the inner power law index, $\gamma$, while holding
constant $\alpha=2$, $\eta=1$, $\tb/\tsis=0.01$, and
$\tpm/\tsis=0.03$. The final frame, with $\gamma=1$, is effectively a
singular isothermal sphere plus SMBH. }
\end{figure*}

Both the cored and broken power law potentials demonstrate these
properties. In both models, the source plane is divided into three
regions by two concentric circular caustics and a degenerate caustic
point at the origin. In fact, as shown in Figure \ref{fig_pos_mag}
with the PM+CIS model, for different values of the source position,
either one or three secondary images may form opposite the position
of the primary image.  For the cases in which we are interested, the
core radius and point mass have little effect on the properties of
the primary image (``A''), but can dramatically change the properties
of the three possible secondary images (``B'', ``C'', and ``D'').

Image B, where it exists, will be a saddle point in the light travel
time and will have negative magnification.  It will lie very close to
the location of the secondary image that would form from a singular
isothermal sphere potential alone, and therefore, provides little
information about the properties of the core of the galaxy or the
SMBH.  Image C will be a maximum of the light travel time and will
have positive magnification.  It will carry information about the
degree of softening of the galaxy's potential, generally becoming
brighter for larger cores and fainter for smaller cores.  Image D
will be another saddle point in the light travel time and again will
have negative magnification.  For most source positions, it will
track the position and magnification of the image that would form
from an isolated point mass potential, thus providing information
about the SMBH.  Image D is present even without images B and C for
source positions with sufficiently large $\beta$.

\begin{figure}
\begin{center}
\includegraphics{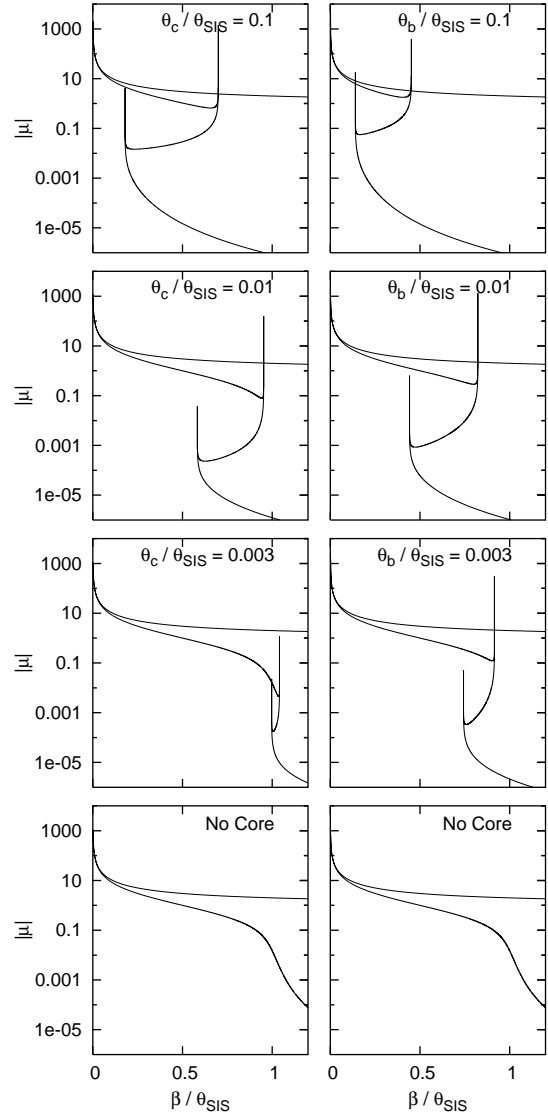}
\end{center}
\caption{\label{fig_radius_sequence} Series of plots illustrating the
progression, from top to bottom, of the magnification curves as the
core radius is reduced. The left column depicts the progression for
the PM+CIS potential. The right column depicts the progression for
the PM+POW potential where $\alpha=2$, $\eta=1$, and $\gamma=0.1$.
For both cases, $\tpm/\tsis=0.03$.  The behavior is qualitatively
similar in both cases.  As the core radius is reduced, the region of
four images moves outward from the lens galaxy and reduces in area
until it disappears at a finite core radius (not shown). }
\end{figure}

\begin{figure*}
\begin{center}
\includegraphics{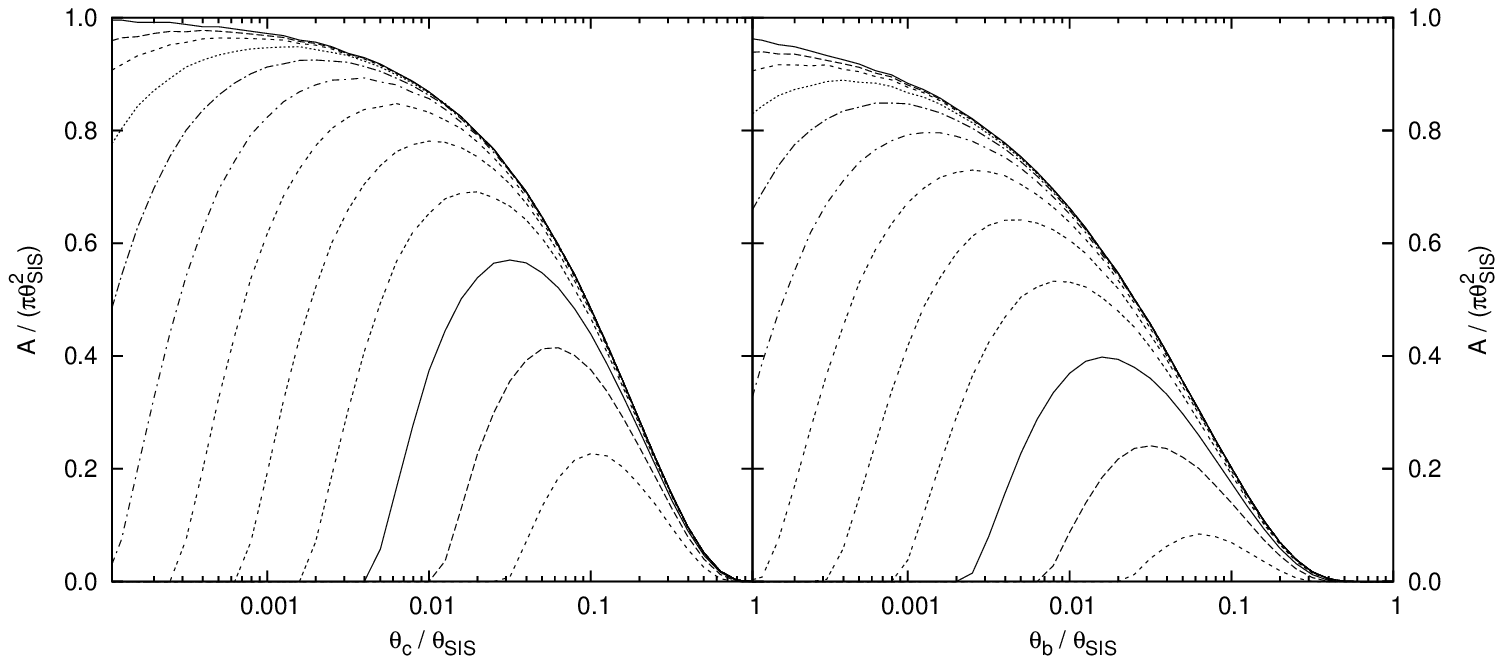}
\end{center}
\caption{\label{cross_section_4_images} Normalized cross-sectional
area, $A/(\pi \tsis^2)$, of the anulus in the source plane where four
images are produced is plotted as a function of core radius. The left
graph is for PM+CIS, the right is for PM+POW.  The curves represent
different values of $\tpm/\tsis$.  Beginning with the lowest curve in
each graph and working up, $\tpm/\tsis = \left \{ 10^{-1}, 10^{-1.2},
10^{-1.4}~... \right \}$. The solid line marking the upper bound of
all curves represents the area of the corresponding three-image
region for a cored isothermal sphere with no SMBH. The reduction of
the curves from the solid line, particularly for small core radii,
represents core image "swallowing" due to the SMBH.}

\end{figure*}

\citet{2003ApJ...582...17K} points out that including SMBHs in
lensing calculations has the affect of ''swallowing'' faint core
images.  Figure \ref{fig_pos_mag} demonstrates that for sufficiently
small impact parameters, where core images would be faintest, the
SMBH does indeed eliminate the core image that would have formed in
its absence. Furthermore, Figures \ref{fig_alpha_gamma_sequence} and
\ref{fig_radius_sequence} demonstrate that, for a given $\tpm/\tsis$,
the size of the region excluding core images is dependent on the
other properties ($\tc$, or $\tb$, $\gamma$, and $\alpha$) of the
inner region of the host galaxy. Referring again to Figure
\ref{fig_radius_sequence}, it is easily seen that for both models,
reducing the core radius pushes the caustics out and closer together
until ultimately they merge (this actually occurs before the core
radius goes to zero, although the exact configuration is not shown in
the plotted sequence). Therefore, smaller core radii decrease the
likelihood of observing four-image lenses as can be seen in Figure
\ref{cross_section_4_images} where we have plotted the area between
the caustics as a function of core radius for multiple curves of
constant $\tpm/\tsis$.  This plot also demonstrates that increasing
$\tpm/\tsis$ also acts to reduce the region of four images.  This
degeneracy accounts for the weakening of constraints on inner
mass-profiles discussed by \citet{2001ApJ...549L..33R}.

\section{NUMERICAL ANALYSIS OF THE MODELS}
\label{sec_numerical}

\subsection{Lensing cross-sections}

As we have seen, the PM+CIS and PM+POW models produce lenses with a
minimum of two images and a maximum of four images, depending on the
source position.  We would like now to extend the analysis and gain
insight into the likelihood of observing these images, particularly
image C and image D.  To do so, we must introduce a parameter which
relates to the observational ability to detect an image.  Let us
assume, for simplicity, that an image can be detected if it produces
a flux that is above a specified fraction of the flux produced by the
primary image:
    \begin{equation}
    \frac{F_{B,C,D}}{F_A} = \frac{\mu_{B,C,D}}{\mu_A} \geq \frac{1}{R},
    \end{equation}
where $R$ is the maximum detectable dynamic range between images and
is the parameter we will use in the following calculations.  The
cross-section for detection is then defined as the area in the source
plane that produces images which meet this requirement, and is in
general characterized by two extreme source positions, $\beta_{\rm
min}$ and $\beta_{\rm max}$, such that the total angular area, $A$,
is given by
    \begin{equation}
    A = \pi \left ( \beta_{\rm max}^2-\beta_{\rm min}^2 \right ).
    \end{equation}
In this expression, $A$, $\beta_{min}$, and $\beta_{max}$ are
implicitly dependent on the core radius, the SMBH Einstein radius,
the galaxy Einstein radius, and the maximum dynamic range.

We are now in a position to calculate the characteristic lensing
cross-sections for the PM+CIS and PM+POW models. We consider three
cross-sections: 1) for detecting two images (A and B, or A and D), 2)
for detecting three images (A, B, and C), and 3) for detecting four
images (A, B, C, and D).  The cross-section for observing three
images is, in fact, the sum of two of the annuli discussed above, one
along the inner caustic and the other along the outer caustic. Due to
the nature of the lensing potential equations, numerical methods were
employed to determine $\beta_{\rm min}$ and $\beta_{\rm max}$ for a
range of model parameters to a precision of $0.002~\tsis$. The
results are discussed below.

\subsubsection{Two Images}

\begin{figure*}
\begin{center}
\includegraphics{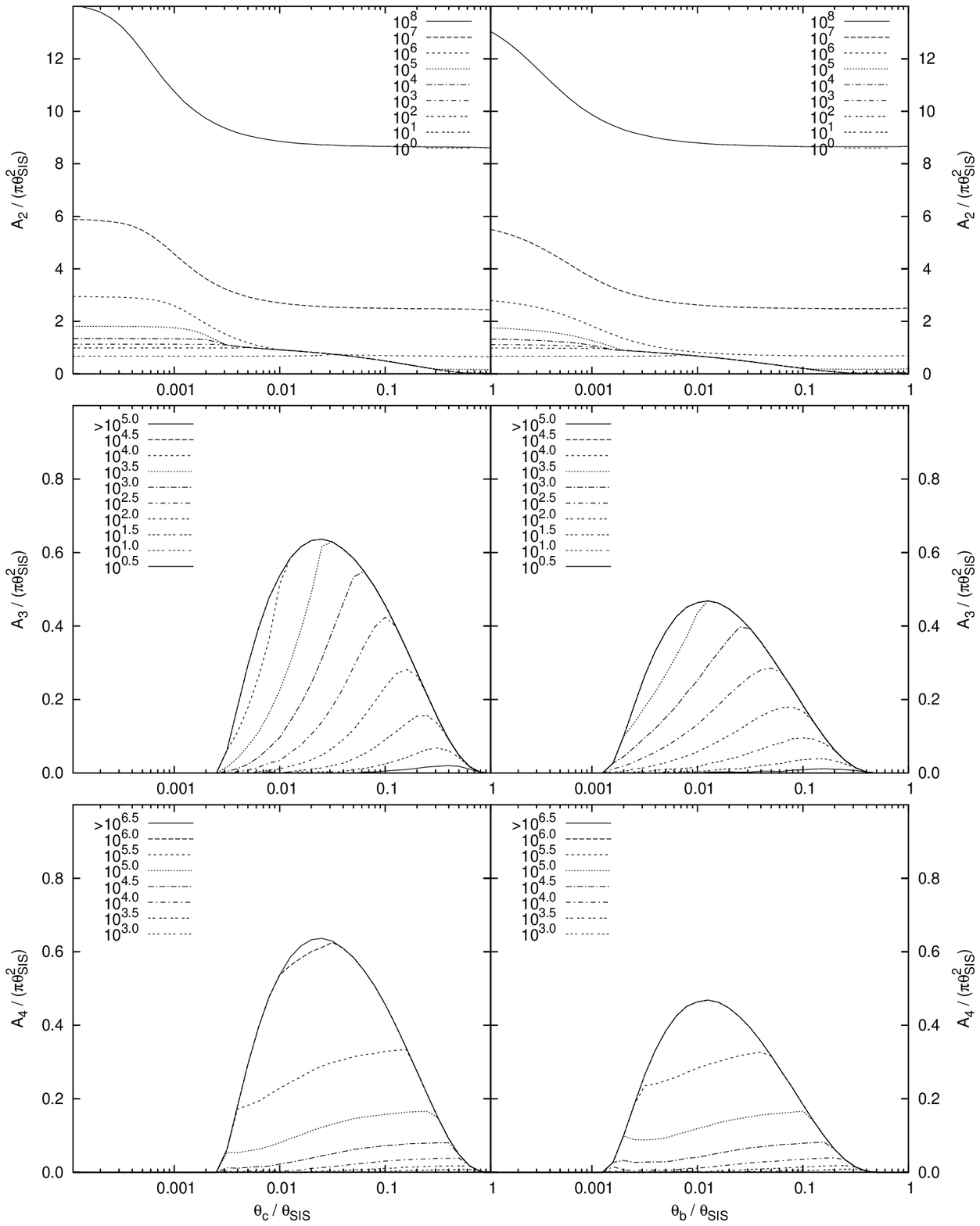}
\end{center}
\caption{\label{fig_cross_section_r} Normalized cross-sectional area,
$A/(\pi \tsis^2)$, in the source plane is plotted as a function of
core radius for $\tpm/\tsis=0.03$. The curves represent different
values of the flux ratio, $R$, ranging from $10^8$ down to $10^0$.
The left column of graphs is for PM+CIS, the right is for PM+POW. The
first row is the cross-section for two images, the second row is the
cross-section for three images, and the third row is the
cross-section for four images.}
\end{figure*}

\begin{figure*}
\begin{center}
\includegraphics{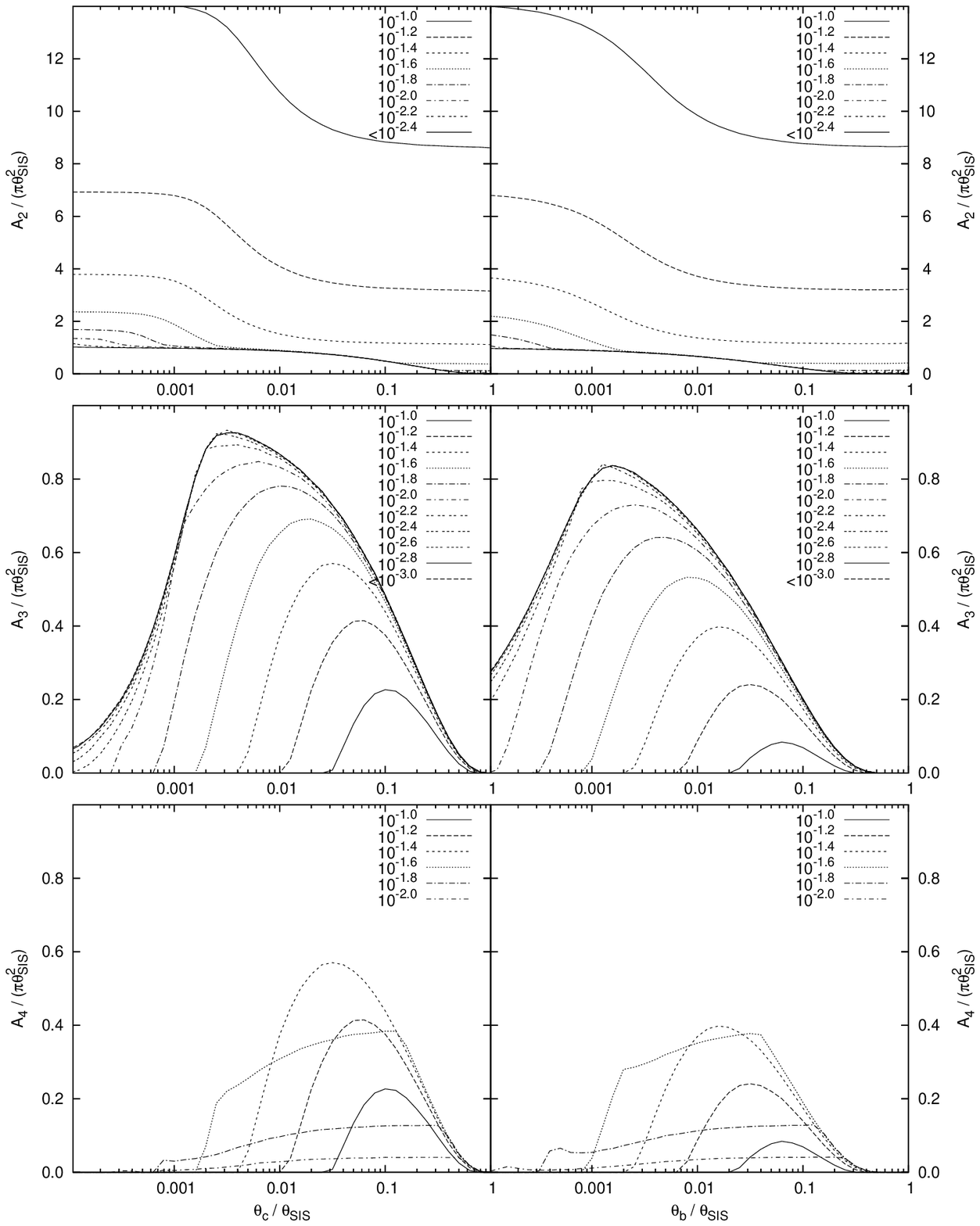}
\end{center}
\caption{\label{fig_cross_section_pm} Normalized cross-sectional
area, $A/(\pi \tsis^2)$, in the source plane is plotted as a function
of core radius for a dynamic range, $R=10^6$. The curves represent
different values of $\tpm/\tsis$ ranging from $10^1$ down to
$10^{-4}$. The left column of graphs is for PM+CIS, the right is for
PM+POW.  The first row is the cross-section for two images, the
second row is the cross-section for 3 images, and the third row is
the cross-section for four images. }
\end{figure*}

It is clear from Figure \ref{fig_radius_sequence} that combining a
point mass and isothermal sphere produces the possibility for a very
large two-image cross-section.  In fact, without a maximum dynamic
range constraint the cross-section would be theoretically unbounded,
and even for finite dynamic range constraints, $\beta_{\rm max}$ may
be well beyond the standard singular isothermal sphere limit of
$\beta_{\rm max}=\tsis$.  To see more quantitatively how dynamic
range affects the cross-section, Figure \ref{fig_cross_section_r}
plots $A_2$ as a function of core radius for several values of $R$,
ranging from $1$ to $10^8$, with $\tpm/\tsis$ fixed at $0.03$. There
are two striking features in this plot.  First, there is a degeneracy
of cross-section curves, around $A_2 / (\pi\tsis^2) \simeq 1$ (the
same feature is visible in Figure \ref{fig_cross_section_pm}). This
is caused by the disparity in magnification between image B and the
image D tail, both of which can contribute to the two-image
cross-section. The disparity reduces as the core radius is reduced
and eventually, for cores less than about $\tc / \tsis \simeq
10^{-2.8}$, disappears because the source plane is no longer divided
into three regions, but only one that always produces two images (see
Figure \ref{fig_radius_sequence}, particularly the bottom row). This
situation eliminates the distinction between our labels ``B'' and
``D'' for images, although, for $\beta \lesssim \tsis$ the secondary
image has properties similar to those of a typical image B and, for
$\beta \gtrsim \tsis$, the magnification drops toward that of a
typical image D.  This smooth decrease in magnification near $\tsis$
for lenses with small core radii is responsible for the second
noticeable feature: cross-sections increase rapidly with dynamic
range for small core radii since there is no step function in
magnification to overcome as is the case for larger core radii.

The SMBH has an effect on the cross-section, as well.  Figure
\ref{fig_cross_section_pm} specifically addresses this dependence,
and from it we see that the two-image cross-section increases with
$\tpm/\tsis$.  This is due to the image D tail increasing in
magnification for larger SMBH mass scales.  The effect is most
dramatic for small core radii.

\subsubsection{Three Images}

For the three-image cross-section, it is also useful to begin by
examining the limit of no dynamic range constraints.  In this case,
the region in the source plane that produces three observable images
is defined completely by the two concentric caustics as discussed at
the end of Section 2 and plotted in Figure
\ref{cross_section_4_images}.  This represents an upper bound for the
three-image cross-section which is plotted in Figures
\ref{fig_cross_section_pm} and \ref{fig_cross_section_r}.  Large
dynamic ranges will saturate to this limit.  As the dynamic range is
lowered, the cross-section reduces with the most rapid decline at
small core radii.  This can be understood from examination of Figure
\ref{fig_radius_sequence}, once again noting that the magnification
of the image C generally decreases as core radius decreases.  It is
no coincidence that the range of core radii that produce a
significant three-image cross-section corresponds to the range where
the two-image cross-section is highly degenerate since both trace the
portion of parameter space that is dominated by the region in the
source plane between the two concentric caustics.

Unlike the two-image cross-section, increasing the SMBH mass scale
decreases the three-image cross-section.  This is another
manifestation of the image ``swallowing'' previously described in
Section \ref{sec_analytic}.

\subsubsection{Four Images}

The four-image cross-section is qualitatively similar to the
three-image cross-section.  However, since the overall magnification
of image D is considerable less, the cross-section curves are
correspondingly reduced as is clear in Figures
\ref{fig_cross_section_pm} and \ref{fig_cross_section_r}.

\subsection{Lensing Statistics}

In this section we derive the optical depth to lensing for a given
source using the calculated cross-sections from above and following
the standard method \citep[][Chapters 11 and 12]{1984ApJ...284....1T,
book_gravitational_lenses}. If we assume that the lensing objects
have a local comoving number density that can be written as
$n_L(\sigma,z)$, then the differential optical depth of a lensing
event occurring at a redshift $z_L$ for a source at $z_S$ is
    \begin{equation}
    d\tau=n_L\left(\sigma,z_L \right ) (1+z_L)^3 A \left ( \sigma,z_L,z_S \right) D_{OL}^2 \left ( z_L \right ) d\sigma c\,dt.
    \end{equation}
Here we constrain $\tpm/\tsis$ according to Equation \ref{eqn_mbh}
and have explicitly written the cross-section, A, as a function of
velocity dispersion, lens redshift, and source redshift to highlight
its dependence on these parameters. It is, of course, still dependent
on the choice of core radius and dynamic range, as well. $D_{OL}$ is
again the angular diameter distance between the observer and lens,
and, in a homogeneous, isotropic universe we have
\citep{astro-ph/9905116}
    \begin{equation}
    c\,dt = \frac{R_0}{1+z_L} \left[\Omega_M\left(1+z_L\right)^3+\Omega_k\left(1+z_L\right)^2+\Omega_\Lambda\right]^{1/2}
    dz_L,
    \end{equation}
where $c$ is the speed of light, $t$ is the lookback time, and $R_0$,
$\Omega_M$, $\Omega_\Lambda$, and $\Omega_k$ are the standard
cosmological parameters with $\Omega_k=1-\Omega_M-\Omega_\Lambda$.
Conceptually, all that remains is to select the form of the density
distribution function.  It is common in gravitational lens studies to
utilize either the Press-Schechter formalism
\citep{1974ApJ...187..425P} or the Schechter luminosity function
\citep{1976ApJ...203..297S} to generate this distribution. We have
chosen to use the Schechter luminosity function combined with the
Faber-Jackson relation and have also assumed no number density
evolution with redshift to cast the density distribution in terms
only of velocity dispersion giving:
    \begin{equation}
    n_L(\sigma)=\gamma
    n_*\left(\frac{\sigma}{\sigma_*}\right)^{\gamma\left(\alpha+1\right)-1}\exp\left[-\left(\frac{\sigma}{\sigma_*}\right)^\gamma\right].
    \end{equation}
We use the same choices of parameters $\alpha$, $\gamma$, and
$\sigma_*$, and abundances, $n_*$, of spiral, SO, and elliptical
galaxy types as \citet{2003MNRAS.343..639O} and have reprinted them
in Table \ref{tab_gal_params}.

The redshift distribution function of extragalactic flat-spectrum
radio sources is localized around $z_s\simeq1$ and has a mean value
of $\langle z_s \rangle = 1.1$ \citep{2003ApJ...594..684M}.  We will
assume that treating all sources as if they were at the mean redshift
is sufficient and take $z_s=1.1$.  We are now prepared to calculated
the optical depth to lensing for each cross-section of interest by
integrating over velocity dispersion and redshift.  Again these
calculations must be carried out numerically because the
cross-sections themselves are not analytical.

%
    \begin{deluxetable}{lccc}
    \tablewidth{3in}
    \tablecaption
    {
        Luminosity function parameters for spiral, SO, and elliptical galaxy types
        \label{tab_gal_params}
    }
    \tablehead{
        \colhead{Parameter} &
        \colhead{Spiral} &
        \colhead{S0} &
        \colhead{Elliptical}
    }
    \startdata
    $\alpha$\dotfill            & $-1.16$                & $-0.54$              & $-0.54$               \\
    $n_{*}$\dotfill             & $1.46\times10^{-2}$    & $0.61\times10^{-2}$  & $0.39\times10^{-2}$   \\
    $\gamma$\dotfill            & $2.6$                  & $4.0$                & $4.0$                 \\
    $\sigma_{*}$\dotfill        & $144$                  & $206$                & $225$                 \\
    \enddata
    \tablerefs
    {
        \citet{2003MNRAS.343..639O}
    }
    \end{deluxetable}

\begin{figure*}
\epsscale{1}
\plotone{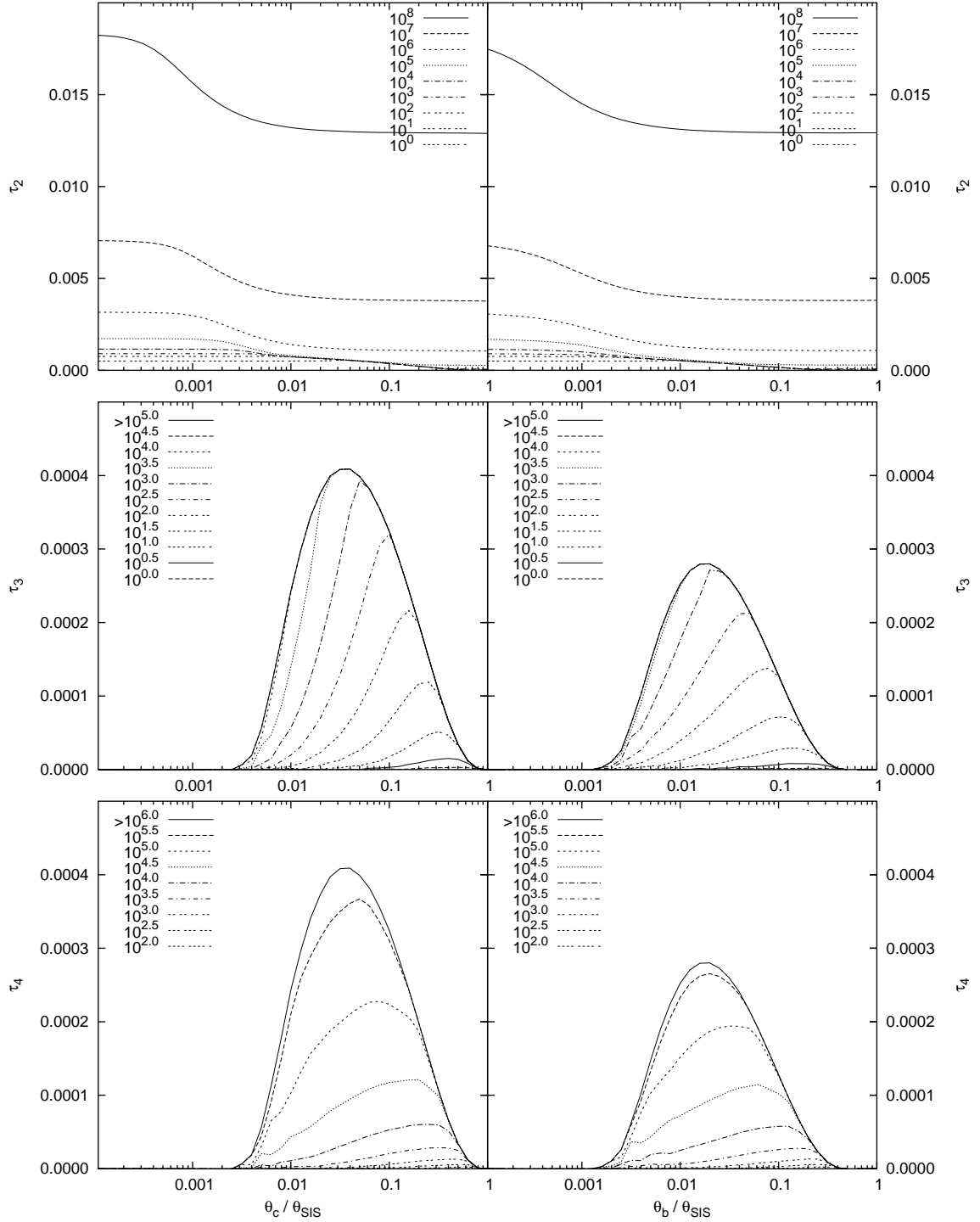}
\caption{\footnotesize \label{fig_optical_depth_z1} Optical depth, $\tau$, for
lensing of a source at redshift, $z_S=1.1$, expressed as a function
of core radius. The curves represent different values of the flux
ratio, $R$, between the brightest image and image of interest.  The
left column of graphs is for PM+CIS, the right is for PM+POW. The
first row is $\tau_2$, the optical depth for observing a lens with at
least two images. The second row is $\tau_3$, the optical depth for
at least three images. The third row is $\tau_4$, the optical depth
for four images. Note that for $\tau_3$, the curves above $R=10^5$
saturate and overlap, with $R=10^{3.5}$ being the first to
significantly break apart.  The same is true for $\tau_4$, however in
this case, the curves above $R=10^{6}$ saturate and $R=10^{5.5}$ is
the first to part significantly.  The saturation is the result of the
finite area of the region between the caustic circles in the source
plane that produces more than two images.  If the source redshift is
increased to $z_S=2.0$, then the optical depth curve for a given $R$
is increased by approximately a factor of 3.  A standard cosmology of
$\Omega_M=0.3$, $\Omega_\Lambda=0.7$, and $H_0=70$km/s/Mpc is used.}
\end{figure*}

%
The results of these calculations are presented in Figure
\ref{fig_optical_depth_z1}. For two-image lensing, we see that the
optical depth to lensing, $\tau$, parallels the cross-section curves
for this case. For high maximum dynamic range, the probability is
quite high, and as the dynamic range is lowered to more typical
values of $R\simeq1000$, the probability approaches the expected
value for lensing by a population of singular isothermal spheres of
$\tau\simeq0.001$ \citep{2003MNRAS.341...13B}, as it must since this
result has been observationally justified.

The optical depth to lensing for three- and four-image cases are also
similar to their cross-sections, with the four image case suffering a
more rapid decline in optical depth as $R$ is decreased. Both
probabilities exhibit significant dependence on the assumed core
radius of the lensing population, indicating that the addition of a
point mass to a softened isothermal lens model, if restricted to the
relationship discussed in Section \ref{sec_typical_values}, would not
hinder using lens statistics to learn about galaxy cores.

\section{EMPIRICAL IMPLICATIONS OF THE MODELS}
\label{sec_empirical}

\subsection{Relevant Facilities and Instruments}

We have seen that if an intervening lens galaxy contains a SMBH, then
an additional faint image is predicted due to the black hole
potential which is not present in unmodified models. In fact, even in
cases when the source position and the lens are too far apart to
create multiple images in the standard way, the SMBH will still
produce a faint image of the background object near the center of the
foreground galaxy. In this section we discuss the applicability of
optical, infrared, and radio facilities to observing these images.

\subsubsection{Optical and Infrared}

Optical observations of these highly demagnified images would be
difficult. In order to obtain a significant cross-section for
detecting image D, we saw in Section \ref{sec_numerical}(and Figure
\ref{fig_cross_section_r}) that demagnification from the primary
image by $10^6$, or 15 magnitudes, must be observable. In the
Hamburg-ESO Survey \citep{2000A&A...358...77W} and the Sloan Digital
Sky Survey (SDSS), the brightest quasars at redshift greater than 0.5
have magnitudes of 15. The demagnification due to the lensing alone
would reduce this to 30th magnitude, and this would be compounded
with the extinction that the image would suffer passing through the
inner regions of the lensing galaxy. Such faint observations are far
beyond the capabilities of current optical telescopes. To observe the
lensing of even the brightest quasars would require 30-meter class or
larger optical telescopes.

Infrared observations would not suffer the severe extinction of the
image passing through the foreground galaxy, however, it would still
be problematic to separate the lens image from foreground galaxy
emission.  The use of spectral lines to distinguish two redshift
components may be able mitigate this problem.

\subsubsection{Radio}

Observations with radio telescopes, while still difficult, are more
feasible. The brightest sources above a redshift of $z=0.5$ in the
NRAO-VLA Sky Survey \citep{1998AJ....115.1693C} and the FIRST sky
survey \citep{1995ApJ...450..559B} have flux densities greater than
20~Jy.  The flux density of image D from such a source would be on
the order of 20~$\mu$Jy, and at radio wavelengths any absorption
through the lensing galaxy should be negligible.  Long integrations
with the VLA and VLBI arrays can already achieve sensitivities of
10~$\mu$Jy per beam \citep{2003NewAR..47..385G} or better, sufficient
to observe faint images from the very brightest radio sources. Future
improvements in radio telescope instrumentation, however, will
dramatically increase the predicted number of observable D images.
Four projects in particular are relevant to making sensitive
observations of lenses and are summarized below.

\paragraph{VLA Expansion Project}
The VLA Expansion Project is already underway to supply the VLA with
modern electronics and also plans to add up to eight additional
antennas to the array at distances out to 350~km from the core
facility. The resulting Expanded VLA (ELVA), scheduled for completion
by 2012, is intended to improve performance in important areas by a
factor of 10. This is expected to extend the detection threshold of
faint sources to 1~$\mu$Jy or less.  With this sensitivity, the EVLA
should also be able to detect the C images for existing radio lenses.

\paragraph{Square Kilometer Array}
The Square Kilometer Array (SKA) is a proposed next generation radio
telescope that offers the most promise for observing lenses with
faint images. The current design goals for the SKA are to achieve a
dynamic range approaching $10^8$ and a detection threshold that could
reach to the 10~nJy level.

\paragraph{Low Frequency Array}
A low frequency radio telescope array has been proposed that would
operate at low frequencies between 30 and 240 MHz.  The design is
intended to achieve a baseline dynamic range of $10^5$, extending to
$10^8$ in some cases, and a detection threshold of less than
1~$\mu$Jy. However, due to the low frequencies used, the expected
best-case angular resolution of about 1~arcsec (at 240~MHz) of the
instrument would prevent some lens configurations from being
resolved.

\paragraph{Atacama Large Millimeter Array}
The Atacama Large Millimeter Array (ALMA) is a millimeter wavelength
telescope under design by Europe and North America.  It will be
located in Llano de Chagnantor, Chile, and consist of at least 64,
12-meter antennas.  The intended resolution is $10$ mas with a
sensitivity approximately 20 times greater than the VLA.
\citep{2003MSAIS...3..292T}

\subsection{Lensing Expectations}

Despite the limitations of current technology, it is not too early to
consider candidate SMBH lensing targets.  The first step is to
estimate the predicted total number of lenses that should be
observable.  We can use the optical depth calculations from the
previous section along with the known flux distribution of radio
sources to accomplish this.  To estimate the number of lenses an
instrument with a limiting sensitivity, $S$, should be able to
detect, we integrate the product of the differential source flux
distribution, $d N / d F$, and the optical depth, $\tau(R,\tc)$. The
appropriate dynamic range, $R$, is given by the ratio between the
sensitivity of the instrument and the flux of the source ($R=F/S$):
    \begin{equation}
    N(S,\tc) = \int \tau \left ( F/S, \tc \right ) \left ( \frac{d N}{d F} \right )
    dF.
    \end{equation}
We are neglecting the dependence on source redshift of both $\tau$
and $d N/ dF$.  Again, we will assume that treating all sources as if
they were at the mean redshift, $z_s=1.1$, is sufficient.  We still
cannot perform this integral fully because we have only evaluated
$\tau$ for specific values of $R$. However, we can achieve an
approximate evaluation by converting the integral to a sum over flux
bins centered on equivalent $R$ values:
    \begin{equation}
    N(S,\tc) \approx \sum_{i}
    \tau(F_i / S,\tc) \left ( \frac{d N}{d F} \right )_{F_i} \Delta
    F_i.
    \end{equation}
The differential number of radio sources as a function of flux in the
range $14.5 \leq F \leq 1500$ $\mu$Jy is \citep{1993ApJ...405..498W}:
    \begin{equation}
    \frac{d N}{d F_{8.44}} = \left ( -4.6 \pm 0.7 \right ) F^{-2.3 \pm
    0.02} \rm ~Jy^{-1} Sr^{-1}.
    \end{equation}
We will use this relation for all flux densities, although it is
likely to overestimate the the number of faint sources ($F \lesssim
1$ nJy) and underestimate the number of bright sources ($F \gtrsim
10$ mJy) as discussed by \citet{1993ApJ...405..498W}. Finally, in
order to perform the sum, we must choose a core radius and an image
of interest.  Two diagnostic core radii regimes are evident from
Section \ref{sec_numerical}.  We begin by considering: 1) the limit
of vanishing cores, and 2) the core radius that produces the maximum
likelihood of observing three- and four-image lenses. The first
regime is approached by setting $\tc/\tsis=10^{-4}$ and performing
the sums described above. Using the PM+CIS model we calculate, for
each image, the number of observable lenses per steradian.

%
    \begin{deluxetable}{rcrrrr}
    \tablewidth{3in}
    \tablecaption
    {
        Predicted number of gravitational lenses per steradian using the
        empirical flux distribution of known radio sources and $\tc/\tsis=10^{-4}$
        \label{tab_emp_4}
    }
    \tablehead{
        \colhead{S ($\mu$Jy)} &
        \colhead{Example} &
        \colhead{$N_{Total}$} &
        \colhead{$N_{3}$} &
        \colhead{$N_{4}$} &
        \colhead{$N_{Outside}$}
    }

    \startdata
    1000 & FIRST Survey & 4 & 0 & 0 & 0 \\
    100 & Typical VLA& 74 & 0 & 0 & 0 \\
    10 & \nodata & 1,500 & 0 & 0 & 2 \\
    1 & EVLA& 30,000 & 0 & 0 & 34 \\
    0.1 & \nodata & 590,000 & 0 & 0 & 670 \\
    0.01 & SKA & 12,000,000 & 0 & 0 & 13,000 \\
    \enddata
    \tablecomments
    {
        This table indicates the predicted number of gravitational lenses per
        steradian for a search with sensitivity, $S$, using the optical depth
        calculations for the PM+CIS potential from the previous section and
        the empirical flux distribution of known radio sources.  The choice
        of core radius is $\tc/\tsis=10^{-4}$ to approximate the limiting case of
        no galaxy cores. $N_{Total}$ is the number of two-image lenses
        predicted, $N_{3}$ is the number of lenses in the subset where a
        third, core image is detectable, and $N_{4}$ is the number of lenses
        in the subset with a fourth, SMBH image.  The last column, labelled
        $N_{Outside}$, is the number of two-image lenses where the fainter
        image is due directly to the SMBH.
    }
    \end{deluxetable}

%
Table \ref{tab_emp_4} illustrates that no lenses with more than two
images are expected.  This is readily explained by noting in Figure
\ref{fig_optical_depth_z1} that the optical depths for three- and
four-image lenses at $\tc/\tsis=10^{-4}$ are negligible for all flux
ratios. The last column in the table, labelled ``$N_{Outside}$'',
indicates the number of two-image lenses where the secondary image
would not be present without the SMBH and, therefore, may still be
able to provide useful information about the SMBH.

The second core radius regime is that of maximum likelihood to
produce three- and four-image lenses.  For the PM+CIS model, this
occurs near $\tc/\tsis=10^{-1.3}$.  The calculated values for this
core radius are listed in Table \ref{tab_emp_1.3}.

%
    \begin{deluxetable}{rcrrrr}
    \tablewidth{3in}
    \tablecaption
    {
        Predicted number of gravitational lenses per steradian using the
        empirical flux distribution of known radio sources and
        $\tc/\tsis=10^{-1.3}$
        \label{tab_emp_1.3}
    }

    \tablehead{
        \colhead{S ($\mu$Jy)} &
        \colhead{Example} &
        \colhead{$N_{Total}$} &
        \colhead{$N_{3}$} &
        \colhead{$N_{4}$} &
        \colhead{$N_{Outside}$}
    }
    \startdata
    1000 & FIRST Survey & 4 & 0 & 0 & 0 \\
    100 & Typical VLA & 73 & 1 & 0 & 0 \\
    10 & \nodata & 1,400 & 18 & 0 & 0 \\
    1 & EVLA & 29,000 & 360 & 1 & 1 \\
    0.1 & \nodata & 580,000 & 7,100 & 13 & 16 \\
    0.01 & SKA & 12,000,000 & 140,000 & 260 & 330 \\
    \enddata
    \tablecomments
    {
        This table indicates the predicted number of gravitational lenses per
        steradian for a search with sensitivity, $S$.  The choice
        of core radius is $\tc/\tsis=10^{-1.3}$, which is the value of $\tc$ that
        produces the largest optical depths for core and SMBH images.
    }
    \end{deluxetable}
%
These values indicate that if typical core radii for galaxies are in
this range, we would expect to find, with an SKA-like instrument, no
more than about 260 lenses per steradian with a fourth, SMBH image
out of about 12 million.  On the other hand, 140,000 lenses per
steradian with a third image bright enough to detect would be
expected.

\subsubsection{Limiting Candidate Targets}

Despite the tremendous number of lenses capable of providing
information about an SMBH that an extremely sensitive search is
predicted to detect, it would be operationally infeasible due to the
required long integrations times to conduct such a campaign without
first reducing the number of candidate targets.

If only sources in a radio catalog are under consideration as
possible targets, the sensitivity limit of the catalog will restrict
the number of two-image lenses and increase the overall ratio of
three- and four-image lenses to the total number of lenses.

There are over 800,000 sources in the FIRST radio catalog
\citep{1997ApJ...475..479W} which has a detection limit of 1 mJy and
covers nearly $10,000$ square degrees of the sky, mostly in the
northern hemisphere. For this analysis, we will assume as we did
earlier, that all the sources in the catalog are at redshift,
$z=1.1$, and follow the same procedure as above. We divide the
catalog into eight flux bins, shown in Table \ref{tab_bins},

%
    \begin{deluxetable}{ccr}
    \tablewidth{3in}
    \tablecaption
    {
        Number of sources in the FIRST radio catalog
        \label{tab_bins}
    }

    \tablehead{
        \colhead{Bin} &
        \colhead{Flux (mJy)} &
        \colhead{$N_{bin}$}
    }
    \startdata
    8 & $10^{3.5}$ & 27 \\
    7 & $10^{3.0}$ & 227 \\
    6 & $10^{2.5}$ & 1,444 \\
    5 & $10^{2.0}$ & 7,418 \\
    4 & $10^{1.5}$ & 26,914 \\
    3 & $10^{1.0}$ & 75,756 \\
    2 & $10^{0.5}$ & 181,766 \\
    1 & 1 \phn\phn & 517,565 \\
    \enddata
    \tablecomments
    {
        The number of sources in the FIRST radio catalog are divided into 8
        bins by flux.  The flux values listed in the table represent the
        floor of each bin.  The number of sources in each bin is used in
        combination with optical depth calculations for the PM+CIS lensing
        potential to calculate the predicted number of lenses contained in
        the catalog (Tables \ref{tab_first_4} and \ref{tab_first_1.3}).
    }
    \end{deluxetable}

%
and perform the sum for the two diagnostic core radii regimes.

The first regime, with $\tc/\tsis=10^{-4}$ and using the PM+CIS
model, predicts the number of observable lenses for the area of the
sky covered by the FIRST survey given in Table \ref{tab_first_4}.

%
    \begin{deluxetable}{rcrrrr}
    \tablewidth{3in}
    \tablecaption
    {
        Predicted number of gravitational lenses in the FIRST catalog using
        $\tc/\tsis=10^{-4}$
        \label{tab_first_4}
    }

    \tablehead{
        \colhead{S ($\mu$Jy)} &
        \colhead{Example} &
        \colhead{$N_{Total}$} &
        \colhead{$N_{3}$} &
        \colhead{$N_{4}$} &
        \colhead{$N_{Outside}$}
    }
    \startdata
    1000 & FIRST Survey & 100 & 0 & 0 & 0 \\
    100 & Typical VLA & 470 & 0 & 0 & 4 \\
    10 & -- & 650 & 0 & 0 & 40 \\
    1 & EVLA & 790 & 0 & 0 & 180 \\
    0.1 & -- & 1100 & 0 & 0 & 460 \\
    0.01 & SKA & 1700 & 0 & 0 & 1100 \\
    \enddata
    \tablecomments
    {
        This table indicates the predicted number of gravitational lenses for a search
        with sensitivity, $S$, in the FIRST catalog of radio sources.  The choice of
        core radius is $\tc/\tsis=10^{-4}$.
    }
    \end{deluxetable}

%
Immediately it is clear that, compared to Table \ref{tab_emp_4}, the
number of total lenses detectable by sensitive instruments has been
drastically reduced. This is because the FIRST survey contains no
sources with flux below $F=1$ mJy and, therefore, the extra
sensitivity has no accessible faint sources on which to operate.

The second core radius regime of interest is where
$\tc/\tsis=10^{-1.3}$.  The calculated values for this core radius
are in Table \ref{tab_first_1.3}.

%
    \begin{deluxetable}{rcrrrr}
    \tablewidth{3in}
    \tablecaption
    {
        Predicted number of gravitational lenses in the FIRST catalog of
        using $\tc/\tsis=10^{-1.3}$
        \label{tab_first_1.3}
    }

    \tablehead{
        \colhead{S ($\mu$Jy)} &
        \colhead{Example} &
        \colhead{$N_{Total}$} &
        \colhead{$N_{3}$} &
        \colhead{$N_{4}$} &
        \colhead{$N_{Outside}$}
    }
    \startdata
    1000 & FIRST Survey & 90 & 2 & 0 & 0 \\
    100 & Typical VLA & 410 & 20 & 0 & 0 \\
    10 & -- & 410 & 130 & 1 & 0 \\
    1 & EVLA & 410 & 320 & 12 & 1 \\
    0.1 & -- & 430 & 320 & 70 & 7 \\
    0.01 & SKA & 580 & 320 & 230 & 86 \\
    \enddata
    \tablecomments
    {
        This table indicates the predicted number of gravitational lenses for a search
        with sensitivity, $S$, in the FIRST catalog of radio sources. The choice of
        core radius is $\tc/\tsis=10^{-1.3}$.
    }
    \end{deluxetable}

%
Although significantly fewer lenses are likely to be contained in the
catalog, the relative number with an observable core image is much
higher (between 50\% and 80\% for EVLA or better levels). In
addition, most of the two-image lenses should already be detectable
with the VLA. It is plausible, under these circumstances, that an
appropriate search guided by the FIRST catalog would be able to
detect and characterize, today, a population of two-image lenses that
will yield a significant number of third and fourth images when
re-observed with future technology.

\begin{figure*}
\epsscale{1}
 \plotone{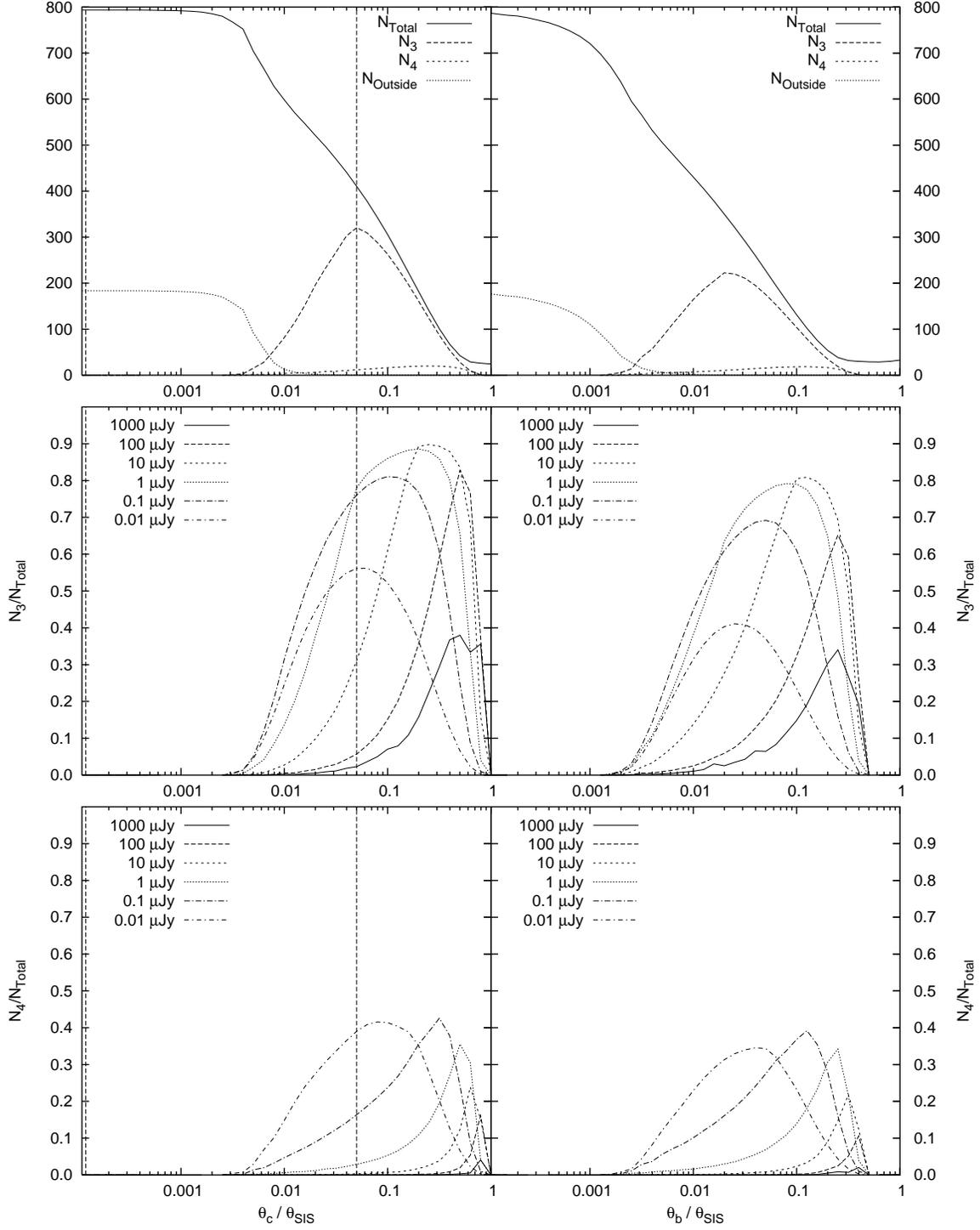} \caption{\label{fig_first_counts}
Number of lenses predicted to be observable in the FIRST radio source
catalog.  The top row is for a search with limiting sensitivity
$S=1\mu$Jy and is plotted as a function of core radius. The curves
are given for the number of lenses with at least two, three, or four
images detectable ($N_2$, $N_3$, and $N_4$, respectively).
$N_{Outside}$ is the number of lenses with two images detectable
where the second image would not be present without the SMBH. In the
middle row, the plots represent the fraction of observable lenses
with detectable third images.  The curves, in this case, represent
different limiting sensitivities from $S=1000\mu$Jy down to
$S=0.01\mu$Jy.  The bottom row of plots is, similarly, the fraction
of observable lenses with detectable 4th images.   For all rows, the
left column of graphs is for PM+CIS, the right is for PM+POW.  The
vertical dashed lines are drawn for the values of $\tc$ used for the
discussion in the text.}
\end{figure*}

The above calculations were performed for two specific core radii of
particular interest, but they can be repeated for all the core radii
we considered in Section \ref{sec_numerical}. This has been done and
is plotted in Figure \ref{fig_first_counts} for S=1 $\mu$Jy.  The
ratios of $N_{3}/N_{Total}$ and $N_4/N_{Total}$ are plotted as well.
For sensitivities better than 10 $\mu$Jy, the $N_{3}/N_{Total}$
increases rapidly with core radius in the region
$\tc/\tsis\geq10^{-2.6}$ and approaches unity for
$\tc/\tsis=10^{-0.6}$ and S=10 $\mu$Jy. Similarly, $N_{4}/N_{Total}$
increases in the same region, although it reaches a maximum of only
approximately $N_{4}/N_{Total} \simeq 0.4$.

Searching for 500 lenses out of a catalog of over 800,000 sources is
still a daunting task.  We suggest that another object catalog may
help reduce the list of candidates even more.  The Sloan Digital Sky
Survey (SDSS) covers 3324 square degrees as of Data Release 2,
overlapping the FIRST survey, and will eventually cover nearly 10,000
square degrees.  It has identified over 88 million objects at optical
wavelengths, many of which are extragalactic sources with estimated
redshifts between $0 \leq z \lesssim 0.5$
\citep{2002AJ....123..485S}. This population of extragalactic objects
makes an ideal foreground of potential lens galaxies for background
radio sources in the FIRST catalog. \citet{2002AJ....124.2364I} have
shown that the error in position between the two surveys is
$\sim0.4$~arcsec per coordinate, and the SDSS has included in its
database a special flag to indicate objects that are spatially
collocated with FIRST radio sources..  If pairs of neighboring
objects within approximately $1~arcsec$ of each other could be found
in the SDSS database where one is marked with the FIRST flag, then
this would represent a sample of targets with an elevated likelihood
of containing radio lenses. If it could be further reduced to pairs
where the radio source is known to be behind the neighboring object,
the likelihood would be higher, yet. Fortuitously, the SDSS aims to
acquire 1 million spectra of galaxies, preferentially targeting
identified optical counterparts of FIRST radio sources. Thus,
redshift information may be available for one or both of the objects
in a pair directly from the database.

A complimentary approach to finding new lens candidates has been
described by \citet{2004AJ....127.1860B}.  They have been able to
identify multiple emission lines at two redshifts in a number of SDSS
spectra, indicating a foreground and background galaxy within the
3~arcsec field of view, and are using these objects in an optical
lens search.

\subsubsection{Galaxy Cores}

It is clear from the preceding estimates that the size of galaxy
cores greatly affects the likelihood of observing core and SMBH
images. It is beneficial, therefore, to consider what is known about
galaxy cores to date.  The structure of the central regions of
galaxies has been studied extensively through surface brightness
measurements and a number of attempts have already been made to
deduce the properties of galaxy cores from gravitational lensing
studies.

Fits to the ``Nuker Law'' surface brightness models, which we used
earlier as the basis of our second lens model, have given the result
that galaxies appear to be of two types, those that are ``cored'' and
those that are not. The cored galaxies are those that are luminous,
slowly rotating ellipticals with boxy isophotes; the non-cored, power
law galaxies are faint, rapidly rotating galaxies with disky
isophotes. There is a fairly tight correlation between break radius
and luminosity.  Large elliptical galaxies dominate the lensing
cross-section, implying that the galaxies that are important for SMBH
lensing are also those that have cores. In fact, it has been
hypothesized that in the hierarchical merger process involved in
galaxy formation, SMBH binary systems will form and spiral toward the
center of the new system, cooling and possibly ejecting starts near
the center.  The SMBHs may, therefore, even be {\it responsible} for
the existence of the cores.

The presence of a core image (image C) in gravitational lenses,
however, is exceedingly rare, with candidate images observed only in
PMN~J1632-0033 \citep{2004Natur.427..613W}, MG~1131+0456
\citep{1993AJ....106.1719C}, and APM~08279+5255
\citep{1999AJ....118.1922I}.  For the latter two systems, it remains
uncertain whether third component is an image or emission from the
lens galaxy.  If we do take the third components in MG~1131+0456 and
APM~08279+5255 as lens images, best fit values of
$\tc/\tsis\simeq0.2$ \citep{1995ApJ...447...62C} and
$\tc/\tsis\simeq0.7$ \citep{2000ApJ...535..561E}, respectively, are
calculated.

The only system with convincing evidence of a core image detection is
PMN~J1632-0033 \citep{2004Natur.427..613W}.  Radio spectra of all
three image components have been acquired and the flux ratios
$F_B/F_A$ and $F_C/F_A$ are nearly constant across the high frequency
measurements indicating the components are likely from the same
source. The only discrepancy is with the lowest frequency measurement
at 1.7 GHz, but as \cite{2004Natur.427..613W} argue, this is where
absorption from passing through the dense part of the lens galaxy
would be strongest. They have, furthermore, been able to place
constraints on several quantities of interest
\citep{2003ApJ...587...80W}.  Using a broken power law model, the
relation between break radius and inner power law slope is
constrained; and using a cored isothermal mass-density model, a core
radius of $\tc \simeq 5.5$ mas or $r_c \simeq 31 h^{-1}$ pc, is
derived.  Compared to the estimated Einstein radius of the lens
galaxy of $\tsis \simeq 0.74$ arcsec \citep{2002AJ....123...10W}, the
core radius corresponds to $\tc / \tsis = 0.0075$, although this is
not directly comparable to our core radius due to the differing
model.

\citet{2003ApJ...582...17K} discusses in detail the implications of
the behavior of the mass-density profile at the core to the
detectability of a third image, and concludes that the paucity of
third images in lens systems is consistent with observed surface
density profiles. \citet{2001ApJ...549L..33R} have studied six double
lens systems in which the third image is not detected. Using the
known detection limits for each of these sources, they employ a
singular mass profile with variable slope to determine how steep the
inner density profile must be to explain the lack of this image. They
find that for these six sources the inner region must be quite steep.
However, it is not clear that the constraints on the inner slope are
relevant in constraining the possible flattening of the mass
distribution that might occur right at the SMBH.

The above attempts have tended to be specific case studies of
individual lenses and therefore it is difficult to get a sense for
the overall probability of any particular core radius.  The results
of our modelling are in a position then, to provide the backdrop for
interpreting these results.  For example the results of a systematic
lens survey with 1 mJy sensitivity could be compared to the ratio of
$N_{3}/N_{Total}$ in Figure \ref{fig_first_counts} to learn about
typical core radii values.

\subsection{Structure in Image D}

To this point we have considered only the overall magnification and
position of image D.  If these images are detected, however, any
observed structure in them will be capable of providing additional
important information about the lens system.  The properties of
observed structure in a lens image should be distinct from the
properties of any radio emission from the foreground galaxy, and
thus, resolved structure in faint emission at the center of a lens
could be used to reduce confusion as to whether an image or
foreground emission had observed. Additionally, structure in image D
is capable of indicating the presence of binary SMBHs in the lensing
galaxy. Both of these are discussed in detail below, but first we
will examine the ability to resolve structure in these images.

\subsubsection{Detecting Structure in Image D}

The flux density of image D, as we have seen, is demagnified
typically by a factor of at least $10^6$ from that of the original
source. Since surface brightness is conserved by lensing, this
corresponds to a reduction in the surface area of the image compared
to the original source.  Therefore, we expect each of the two spatial
dimensions on the sky to be reduced by a factor of order
$\sqrt{10^6}=10^3$.

VLBI measurements achieve angular resolutions approaching $0.1$ mas.
Thus a radio source with $1$ arcsec scale structure would produce an
image D with $\sim1$ mas scale structure, which should easily be
distinguished by the VLBI.

\subsubsection{Foreground Galaxy Emission}

Observing a faint signal near the center of a lens galaxy does not
provide sufficient evidence to prove that image D has been detected.
It is possible that the signal may be explained by a lens model
without a SMBH or that the emission may be from the lens galaxy
itself.

In the former case, an analysis for plausible lens models of the
probability of producing a faint image close to the center of the
lens galaxy should typically be sufficient to remove confusion.  For
example, a standard singular isothermal sphere model produces a faint
secondary image in rare cases where the source impact parameter is
just inside the Einstein radius.  This image, in principle, could be
confused with the image D tail of the models we have discussed above.
However, the probability of such a configuration in the SIS model is
extremely low compared to that for the PM+CIS and PM+POW models.

In the latter case, structure in the signal could be used to rule out
the possibility that the detected signal is emission from the
foreground galaxy. Gravitational lenses produce a characteristic
arcing in images that are resolved. In addition, structure in
secondary images should be correlated with structure in the primary
image.

\begin{figure}
\begin{center}
\includegraphics[width=3in]{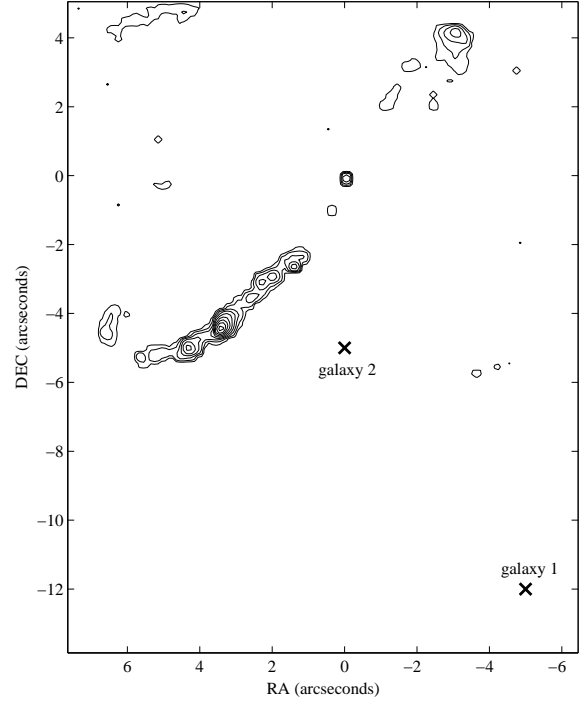}
\end{center}
\caption{\label{fig_structure_s} Contour plot of the radio source
3C9.  The total flux is approximately 190 mJy.  The location of two
nearby galaxies are also indicated.  Galaxy 1 is about 10 arcsec away
from the source, and galaxy 2 is about 3 arcsec away. }
\end{figure}

\begin{figure}
\begin{center}
\includegraphics[width=3in]{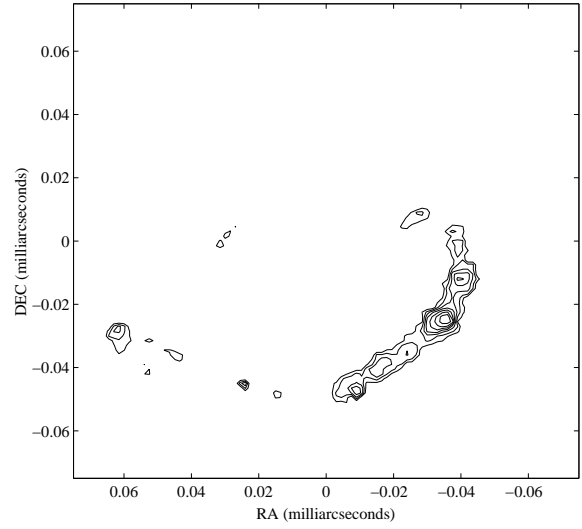}
\end{center}
\caption{\label{fig_structure_l1} Simulated lens image of 3C9
produced by a $10^9 M_{\odot}$ SMBH at the center of galaxy 1.  The
position of the black hole in this plot is the upper left corner. The
total flux of the image is approximately 0.03 nJy. }
\end{figure}

\begin{figure}
\begin{center}
\includegraphics[width=3in]{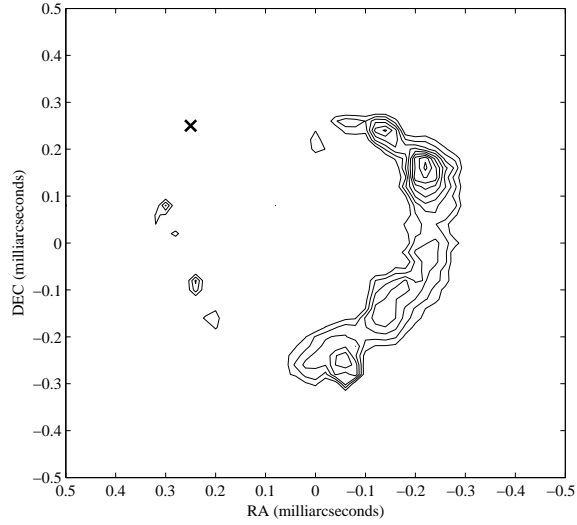}
\end{center}
\caption{\label{fig_structure_l2} Simulated lens image of 3C9
produced by a $10^9 M_{\odot}$ SMBH at the center of galaxy 2.  The
position of the black hole is marked by the cross.  The total flux of
the image is approximately $3$ nJy. }
\end{figure}

As an illustrative example of the lensing that we expect to occur due
to a SMBH within a galaxy, we consider a radio-jet QSO that is known
to have nearby foreground galaxies.  The radio and optical
observations of 3C9 are discussed in detail by
\citet{1991ApJ...367L...1K}.  This source lies at a redshift of
$z_S=2.012$, has total flux of approximately 190~mJy, and its radio
jets are approximately 10~arcsec across. A radio map of the source is
shown in Figure \ref{fig_structure_s}, which was constructed from
data provided by P. Kronberg.  The especially interesting feature of
3C9 is that there are two foreground galaxies near the jets. One
(galaxy 1) lies approximately 10~arcsec from the jets, has an R-band
magnitude of 19.5, and has a measured redshift of $z_L=0.254$. The
other (galaxy 2) lies only about 3~arcsec from the jets, has a R-band
magnitude of 23.0, and has a less well determined redshift. We will
assume that the two galaxies lie at the same redshift ($z_L=0.254$).
Using a simple ray tracing algorithm, we simulated the lensing that
would occur from a SMBH in the center of each galaxy.  Since the
foreground galaxies are far from the radio source, and in such cases
the properties of image D are determined almost entirely by the SMBH
without influence from the host galaxy potential, we are able to work
in the simplifying approximation of treating the SMBH as an isolated
point mass. Figure \ref{fig_structure_l1} shows the image that would
result from a $10^9~M_{\odot}$ SMBH in the center of galaxy 1.  The
total flux from this image is approximately 0.03~nJy.  A SMBH with
the same mass in galaxy 2 would produce a much brighter image with
total flux about 3~nJy.  Figure \ref{fig_structure_l2} shows this
case.

These faint images of 3C9 would not be detectable, even with the
anticipated sensitivity of the SKA, and are presented merely as
examples of how structure could be identified.  3C9, therefore, is
not a realistic candidate for future SMBH lensing studies.

\subsubsection{Mergers and Binary SMBHs}
\label{sec_binary}

An interesting question arising from the hierarchical formation of
SMBHs is how long the black holes of two merging galaxies take to
coalesce. Indeed, if the time scale is long enough we would expect
some galaxies to contain binary SMBHs. \citet{2002MNRAS.331..935Y}
has carried out theoretical studies that determine how these merger
time scales vary according to the type of host galaxy, its velocity
dispersion, and the mass ratios of the SMBHs involved. The
``surviving'' binary systems (those whose merger time is larger than
the Hubble time) are found to occur predominantly in galaxies that
are not highly flattened and that have dispersion velocities greater
than 90~km/s. In addition, SMBH pairs with equal masses are more
likely to survive.

The separations of most of the remaining binary systems range from
0.01 to 10~pc. From Equation \ref{eqn_tpm}, we calculate that for
SMBHs with masses between $10^6$ and $10^9 M_{\odot}$, typical values
of $\tpm$ correspond to physical distances between 10 and 400~pc. The
binary separations, therefore, may be significant fractions of the
SMBH Einstein radii and binaries may be distinguishable in some
lenses.  Binary SMBHs with the widest anticipated separations may be
sufficient to alter not only properties of image D, but also image C
(and therefore the lensing statistics of these images), because for
small source impact parameters image C is predicted to form typically
within about 10~pc of the center of the lens (see Figures
\ref{fig_pos_mag} and \ref{fig_rc_to_tc}). For large source impact
parameters, however, image C does not exist and only image D will be
affected.

\begin{figure}
\begin{center}
\includegraphics[width=3in]{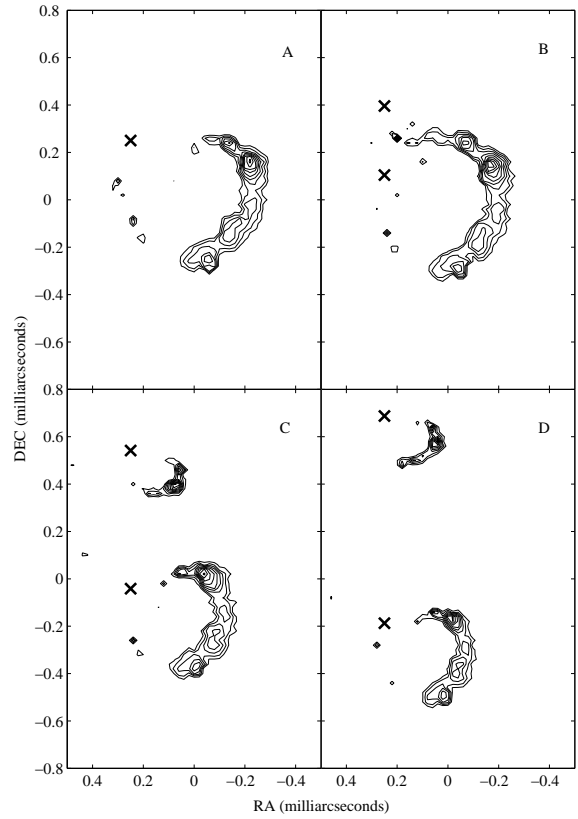}
\end{center}
\caption{\label{fig_structure_binary} Lensed images of 3C9 produced
by a binary SMBH system in galaxy 2.  Panel A shows the image
produced by a single $10^9 M_{\odot}$ SMBH at the position of the
cross.  Panels B, C, and D are the images produced by two $5 \times
10^8 M_{\odot}$ SMBHs separated by $0.01\tpm$ (2.5 pc), $0.02\tpm$
(5.0 pc), and $0.03\tpm$ (7.5 pc), respectively.  The total flux for
each panel, from A to D, is: 3.1 nJy, 4.0 nJy, 3.5 nJy, and 2.3 nJy.
}
\end{figure}

Figure \ref{fig_structure_binary} demonstrates the results of
considering two $5 \times 10^8 M_{\odot}$ SMBHs in galaxy 2 of the
3C9 example considered above.  We calculate for these SMBHs that the
Einstein radii correspond to 250~pc (using $\Omega_M=0.3$,
$\Omega_{\Lambda}=0.7$, and $H_0=70$~km/s). It is evident that, even
for separations of a few parsecs, noticeable effects are visible in
this scenario.  A survey of a large number of these faint lenses
would provide insight not only into the distribution of the masses of
SMBHs, but also whether the SMBHs occur as individual or binary
systems and their correlation to the morphological type of the host
galaxy.

It is also evident from theoretical simulations that, either through
mergers or initial conditions, individual SMBHs may not be located
precisely at the center of their host galaxies. If a SMBH has formed
at a distance greater than approximately 100~pc (or more, depending
on its mass) from the center of its host galaxy, \citet[Figure
3]{2001PhRvD..64d3504U} have predicted that dynamical friction is
insufficient to produce spiral-in timescales shorter than the Hubble
time.  Additionally, Brownian motion may cause a SMBH or binary pair
to wander around the nucleus of a galaxy within a radius of order
0.01~pc \citep{2001ApJ...556..245M}.  Observational evidence, on the
other hand, suggests that in nearby galaxies SMBHs are within 0.1~pc
of galaxy centers \citep[Section 1]{2001ApJ...563...34M}.  As was the
case for binary systems, non-centrally located SMBHs may affect the
magnifications curves for small source impact parameters, and thus
the lensing statistics for images C and D. However, they are not
likely to significantly alter the properties of systems with large
source impact parameters since, in this limit, the properties of
image D approach those of an image from an isolated SMBH (see
Appendix \ref{sec_offcenter} for a justification). Therefore, an
off-center SMBH should not complicate detection or identification of
a SMBH image in a two-image system with large source impact
parameter.

\section{CONCLUSION}
\label{sec_conclusion}

Observational measurements of the relationship between SMBHs and the
properties of their host galaxies are an important method for probing
theoretical hierarchical growth models. Gravitational lensing is a
unique mechanism for acquiring the this information in systems at
cosmologically significant redshifts.

We have reviewed the calculations required to transform historically
popular galactic lens models in order to account for the presence of
SMBHs.  The modified models, PM+CIS and PM+POW, exhibit an inner disk
in the source plane where only two bright images are produced,
surrounded by an anulus where two additional faint images are added
near the center of the lens for a total of four images.  For all
source positions outside this region, the models predict the presence
of one primary image and one highly demagnified image deep in the
center of the lens which traces very nearly the secondary image of an
isolated SMBH.  The relative cross-sections of these three regions
are very sensitive to galaxy core sizes and SMBH masses.  We used
these characteristic signatures to demonstrate the origin of the SMBH
lensing effects reported previously in \citet{2001ApJ...549L..33R},
\citet{2003ApJ...582...17K}, and \citet{2001MNRAS.323..301M}.

Determination of lensing cross-sections over a range of parameter
space and subsequent optical depth calculations revealed that, even
for the most favorable lens configurations, observations of the
highly demagnified fourth images will remain elusive until the next
generation of radio telescopes are in use.  Two scenarios, however,
offer more immediate results.  Since core images are detectable now,
if galaxies tend to have large cores, deep VLBI observations (or
EVLA) should make accessible a sufficiently large population of core
images to provide meaningful statistics. On the other hand, if
galaxies tend to have very small cores, then two-image lenses where
the secondary image is the highly demagnified SMBH image already may
be observable. For any scenario, there is strong evidence suggesting
that a targeted search today for new lenses, utilizing sky surveys
such as FIRST to limit the number of candidate objects, should yield
a precursory sample of two-image lenses that will, with improved
technology, provide a significant number of third and fourth images.
And finally, if these images are detected they may even reveal binary
SMBH systems.

\vspace{12pt} We would like to thank Philipp Kronberg for providing
the radio map of 3C9.  Support for this work was provided by the NSF
through grant AST-0071181.


\setcounter{equation}{0}  
\setcounter{section}{0}
\renewcommand{\thesection}{\Alph{section}}
\renewcommand{\theequation}{\Alph{section}-\arabic{equation}}

\section{MORE ON NON-CIRCULARLY SYMMETRIC POTENTIALS}
\label{sec_appendix}

Many observed gravitational lenses do not produce simply two bright
images as a circularly symmetric potential would predict. Instead,
four or more bright images are often observed. Two effective and
common explanations are: 1) that the lensing galaxies are indeed not
circularly symmetric but rather elliptical, and 2) that the lensing
galaxies lie in an external gravitational shear.  In this section we
provide a cursory analysis of how ellipticity and shear will modify
the results of the previous sections. In addition, we treat the case
of an off-center SMBH in the large source impact parameter limit.

We have seen in the case of circularly symmetric potentials that, for
large impact parameters ($\beta$), the magnification of image D is
essentially independent of the lensing properties of the galaxy that
contains the image.  We will show that neither ellipticity nor shear,
will significantly alter the magnification of this image or the
primary image, and therefore, the optical depths and statistics
pertaining to observing these images calculated for the circularly
symmetric potentials do not need to be altered.  The properties of
any other faint images, however, will be complicated by the addition
of the more complex potentials and our previous calculations may not
necessarily hold.  Similarly, an off-center SMBH is shown not to
alter the magnification of image D in the large source impact
parameter limit, but other configurations and other images may be
affected.
\subsection{Ellipticity}
It has been demonstrated \citep{2002ApJ...575...68E} that an
elliptical lensing potential derived from a cored isothermal sphere
increases the maximum number of images produced from three to five.
For our PM+CIS potential, considering ellipticity does not actually
increase the maximum number of images, but rather complicates the
caustics in the source plane considerably
\citep{2001MNRAS.323..301M}. We expect that the magnification of the
SMBH image, where it exists, and of the primary image will be
affected little by this change, but to demonstrate this
quantitatively we must return to the lens equation. Recalling from
Equation \ref{eqn_lens}, but this time in two dimensions, the lens
equation is:
    \begin{equation}
    \boldsymbol\beta = \boldsymbol\theta - \boldsymbol\nabla \Psi.
    \end{equation}
In order to describe a galaxy with an elliptical density
distribution, we will modify the circularly symmetric cored
isothermal sphere lensing potential to \citep[][Section
3.5.1]{astro-ph/9606001}:
\begin{equation}
    \Psi_{elliptical}(\tx,\ty) = \tsis \sqrt{\theta_{\rm C}^2
    +(1-\epsilon)\tx^2+(1+\epsilon)\ty^2} .
\end{equation}
In this equation the variables $\tx$ and $\ty$ are the angular
distances along the principal axes of the ellipse, and $\epsilon$ is
the ellipticity of the potential. The cored elliptical lensing
potential is not entirely realistic since, for large ellipticity, it
results in dumb-bell shaped mass-density contours.  For small
ellipticity, however, the contours are generally elliptical and the
potential is quite acceptable. Combining this with the point mass
potential yields the complete lensing potential for an elliptical
galaxy with a SMBH:
\begin{equation}
    \Psi(\tx,\ty) = \tsis \sqrt{\theta_{\rm C}^2
    +(1-\epsilon)\tx^2+(1+\epsilon)\ty^2} + \tpm^2 \ln
    \sqrt{\tx^2 + \ty^2} .
\end{equation}
Using the lens equation, we find for the components of $\beta$:
    \begin{eqnarray}
    \beta_x = \tx \left [ 1 - \frac{\tsis(1 - \epsilon)}{\sqrt{\theta_{\rm C}^2 + (1-\epsilon)\tx^2 + (1+\epsilon)\ty^2}} - \frac{\tpm^2}{\tx^2+\ty^2} \right ] \\
    \beta_y = \ty \left [ 1 - \frac{\tsis(1 + \epsilon)}{\sqrt{\theta_{\rm C}^2 + (1-\epsilon)\tx^2 + (1+\epsilon)\ty^2}} - \frac{\tpm^2}{\tx^2+\ty^2} \right ]
    \end{eqnarray}
Since we are interested only in magnification of images for the
present purposes, we can forgo completely solving the lens equation.
For non-circularly symmetric potentials, the magnification of any
image is given by \citep[][Equation 5.21]{book_gravitational_lenses}:
    \begin{equation}
    \mu^{-1} = \det \mathcal{A},
    \end{equation}
where $\mathcal{A}$ is the Jacobian matrix which describes the lens
properties, and is:
    \begin{equation}
    \mathcal{A} = \frac{\partial \boldsymbol \beta}{\partial \boldsymbol \theta} =  \left ( \begin{array}{cc}
    \partial_x \beta_x & \partial_y \beta_x \\
    \partial_x \beta_y & \partial_y \beta_y
    \end{array} \right ).
    \end{equation}
Unfortunately, writing out $\mathcal{A}$ and calculating its
determinant is not particularly illuminating as it is difficult to
see past the algebra.  As a simplification, we consider the limiting
cases of $\theta \ll (\tpm \lesssim \tc)$ and the opposite extreme
$\theta \gg (\tpm \lesssim \tc)$.  The former happens to correspond
to the region of the image plane where SMBH images are expected to be
observed because, from our earlier analysis, we might reasonably
assume $\tc/\tsis\simeq10^{-1}$, $\tpm/\tsis\simeq10^{-2}$, and for
SMBH images, $\theta/\tsis\simeq10^{-3}$.  The latter corresponds to
the region where the primary image is observed since $\theta/\tsis
\gtrsim 1$ for such images and the other parameters, of course, do
not change.

For the case of the inner region of the image plane, where the SMBH
image is expected to form, the components of $\beta$ reduce to:
    \begin{eqnarray}
    \lim_{\theta \ll \tpm} \beta_x = \tx \left [ - \frac{\tpm^2}{\tx^2+\ty^2} \right ] \\
    \lim_{\theta \ll \tpm} \beta_y = \ty \left [ - \frac{\tpm^2}{\tx^2+\ty^2} \right ]
    \end{eqnarray}
and thus the magnification for an image in this region is given by:
    \begin{equation}
    \mu^{-1}_{SMBH} = \det \mathcal{A} \approx - \frac{\tpm^4}{\left ( \tx^2 +
    \ty^2 \right )^2 } = - \frac{\tpm^4}{\theta^4 } .
    \end{equation}
The magnification of these images would not, therefore, be affected
by the ellipticity of the lens galaxy.  It should be noted, however,
that this approximation breaks down in this region if $\tc/\tsis$ is
sufficiently small ($\simeq\theta^2/\tpm^2$).

For the case of the outer region of the image plane, where a primary
image forms, the components of $\beta$ reduce to another pair of
relatively simple expressions:
    \begin{eqnarray}
    \lim_{\theta \gg \tc} \beta_x = \tx \left [ 1 - \frac{\tsis(1 - \epsilon)}{\sqrt{(1-\epsilon)\tx^2 + (1+\epsilon)\ty^2}} \right ] \\
    \lim_{\theta \gg \tc} \beta_y = \ty \left [ 1 - \frac{\tsis(1 + \epsilon)}{\sqrt{(1-\epsilon)\tx^2 + (1+\epsilon)\ty^2}} \right
    ].
    \end{eqnarray}
Along the axes, $\hat \tx$ and $\hat \ty$, the magnification in this
region is given by:
    \begin{equation}
    \mu^{-1}_{Primary} = \det \mathcal{A} \approx
        \left \{
        \begin{array}{cc}
            1 - \frac{\tsis \left ( 1 + \epsilon \right )}{\tx \sqrt{1 - \epsilon}} & \mbox{when $\ty = 0$}, \\
            1 - \frac{\tsis \left ( 1 - \epsilon \right )}{\ty \sqrt{1 + \epsilon}} & \mbox{when $\tx =
            0$}.
        \end{array}
        \right.
    \end{equation}
We can consider that, in effect, the addition of ellipticity simply
modifies the value of $\tsis$ depending on the position of the
primary image. Since the value of $\epsilon$ is at the very greatest
1, and is usually much smaller, we can ignore this contribution as a
minor correction.

\subsection{Gravitational Shear}

\begin{figure}
\begin{center}
\includegraphics{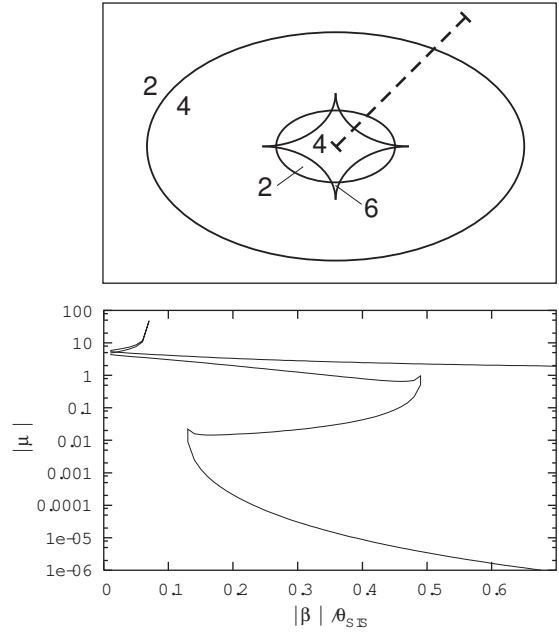}
\end{center}
\caption{\label{fig_shear_caustics} Stylized schematic of the
caustics in the source plane for the sheared PM+CIS potential. The
number of images produced in each region is given.  The relative
sizes of the caustic curves are dependent on the choice of parameters
and the region producing six images is typically not present for
small core radii. The dashed line indicates the approximate path of
source positions for which the image magnifications are plotted in
the bottom panel. In the bottom plot, the source position is given in
units of the singular isothermal sphere Einstein radius, $\tsis$, and
we use $\tc/\tsis=0.1$, $\tpm/\tsis=0.03$, and $\gamma=0.1$.}
\end{figure}

The lensing galaxy might be residing, on the other hand, in the
presence of an external gravitational shear. It has been demonstrated
\citep{2002ApJ...575...68E} that the addition of an external shear to
a cored isothermal sphere potential increases the maximum number of
images produced from three to five, just as did the consideration of
ellipticity.  For our PM+CIS potential, the maximum number of images
increases again to six, however four images is a more probable
outcome for most configurations, as shown in Figure
\ref{fig_shear_caustics}. Again we expect that the SMBH image and the
primary image will be affected little by this change. Returning to
the two-dimensional lens equation, we begin by characterizing an
external shear with \citep[][Section 3.5.2]{astro-ph/9606001}:
    \begin{equation}
    \Psi_{\rm shear} (\tx,\ty) = \frac{\gamma}{2}(\tx^2-\ty^2).
    \end{equation}
Here $\gamma$ is the magnitude of the shear and, as in the case of
the ellipticity, the angular distances $\tx$ and $\ty$ measure along
the axes of the shear.  In order to consider how this affects the
images, we begin by adding this contribution to the potential for a
PM+CIS giving:
    \begin{equation}
    \Psi (\tx,\ty) = \tsis \sqrt{\tx^2 + \ty^2 + \tc^2} + \tpm^2\ln \left ( \sqrt{\tx^2+\ty^2} \right ) +
    \frac{\gamma}{2}(\tx^2-\ty^2) .
    \end{equation}
From the lens equation we again can calculate the components of $\vec
\beta$:
    \begin{eqnarray}
    \beta_x = \tx \left ( 1 - \frac{\tsis}{\sqrt{\tx^2 + \ty^2 + \tc^2}} -  \frac{\tpm^2}{\tx^2 + \ty^2} - \gamma \right ) \\
    \beta_y =  \ty \left ( 1 - \frac{\tsis}{\sqrt{\tx^2 + \ty^2 + \tc^2}} - \frac{\tpm^2}{\tx^2 + \ty^2} + \gamma \right )
    \end{eqnarray}
and for convenience we will define an angle $\omega$ which measures
the position of the image with respect to the positive $\hat\tx$ axis
such that:
    \begin{eqnarray}
    \tx = \theta\cos\omega \\
    \ty = \theta\sin\omega.
    \end{eqnarray}
%
%
%
%
%
%
Calculating the determinant of the Jacobian matrix $\mathcal{A}$ and
grouping the terms by powers of $\gamma$ yields, after some
simplification,
    \begin{eqnarray}
    \nonumber \det\mathcal{A} & = & \left [ 1 - \frac{\tpm^4}{\theta^4} + \frac{\tsis^2 \tc^2}{\left
    ( \tc^2 + \theta^2 \right )^2 } - \tsis \frac{ 2 \tc^2 + \tpm^2 +
    \theta^2 }{ \left ( \tc^2 + \theta^2 \right )^{3/2}} \right ] +  \\
    & & \mbox{} + \gamma \cos 2 \omega  \left [ 2 \frac{\tpm^2}{\theta^2} +
    \frac{\tsis \theta^2}{ \left ( \tc^2 + \theta^2 \right )^{3/2}} \right ] -
    \gamma^2.
    \end{eqnarray}
Following the same approximations here as we did for the elliptical
case, we consider the two regions of the image plane where $\theta
\ll (\tpm \lesssim \tc)$ and $\theta \gg (\tpm \lesssim \tc)$.  For
the inner region of the image plane where an SMBH image is expected
to be observed, and $\theta \ll (\tpm \lesssim \tc)$, we substitute,
as we did before, the values of $\tc/\tsis\simeq10^{-1}$,
$\tpm/\tsis\simeq10^{-2}$, and $\theta/\tsis\simeq10^{-3}$.  In this
limit, the above expression reduces to:
    \begin{equation}
    \lim_{\theta \ll \tpm} \left ( \det \mathcal{A} \right ) \rightarrow - \frac{\tpm^4}{\theta^4}  + 2 \gamma \cos 2 \omega \frac{\tpm^2}{\theta^2} -
    \gamma^2 .
    \end{equation}
Typical values of the shear parameter are generally of order $\gamma
\lesssim 0.1$ \citep{2003ApJ...598L..71C}. Thus, it is sufficient to
keep only the leading term of the expression for the magnification:
    \begin{equation}
    \mu_{SMBH}^{-1} \approx - \frac{\tpm^4}{\theta^4} .
    \end{equation}
This is identical to the elliptical result for this region and we
conclude that shear will not substantially alter the magnification of
an SMBH image for such a lens configuration.

For the case of the outer region of the image plane, where a primary
image forms, and $\theta \gg (\tpm \lesssim \tc)$, we proceed by
repeating the above simplification, but with typical values of
$\theta/\tsis\simeq1$. The determinate of $\mathcal{A}$ is then
reduced to:
    \begin{equation}
    \lim_{\theta \gg \tc} \left ( \det \mathcal{A} \right ) \rightarrow 1
    - \frac{\tsis}{\theta} + \gamma \cos 2 \omega  \left [
    \frac{\tsis}{\theta} \right ] - \gamma^2
    \end{equation}
and using $\gamma \lesssim 0.1$,
    \begin{equation}
    \mu_{Primary}^{-1} \approx 1 - \frac{\tsis}{\theta} \left [ 1 -  \gamma \cos 2 \omega
    \right ] .
    \end{equation}
The angular dependence in this equation makes sense, because symmetry
dictates that two points separated by 180 degrees must have the same
magnification.  As was the case for the elliptical potential, this
can be considered to be a small correction to $\tsis$ and, therefore,
the presence of an external shear has not altered the ratio of the
magnification of a SMBH image to a primary image in these limiting
cases.  Figure \ref{fig_shear_caustics} plots the image
magnifications for a representative system with $\gamma=0.1$ as
function of source position for a path extending away from the center
of the lens. Comparison with the top-left plot in Figure
\ref{fig_radius_sequence}, which has no shear but is otherwise the
same configuration, demonstrates the minimal effect of shear on image
magnifications for both core images in this case.
\subsection{Off-Center SMBH}
\label{sec_offcenter}
Another possible departure from circular symmetry, raised in Section
\ref{sec_binary}, is that the SMBH may not be located in the center
of the host galaxy.  Returning to the lensing potential given in
Equation \ref{eqn_softlenspot}, we can explore the consequences of
such an occurrence by inserting an offset, $\boldsymbol\theta_0$,
between the point mass and the galaxy potential,
    \begin{equation}
    \Psi = \tsis \sqrt{|\boldsymbol\theta|^2 + \tc^2} + \tpm^2 \ln | \boldsymbol\theta - \boldsymbol\theta_0 |.
    \end{equation}
This yields the lens equation,
    \begin{equation}
    \boldsymbol\beta = \boldsymbol\theta - \tsis \frac{\boldsymbol\theta}{\sqrt{|\boldsymbol\theta^2|+\tc^2}} - \tpm^2 \frac{\boldsymbol\theta - \boldsymbol\theta_0}{|\boldsymbol\theta - \boldsymbol\theta_0|^2} .
    \end{equation}
Following a similar approximation here as we did for the elliptical
and shear cases, we consider the region of the image plane which
corresponds to large source impact parameters.  In this limit, we
expect that an image will form near the point mass, and thus
$|\boldsymbol\theta - \boldsymbol\theta_0|$ will be small. The last
term in the lens equation, therefore, will dominate and we can
simplify the lens equation to,
    \begin{equation}
    \boldsymbol\beta \approx - \tpm^2\frac{\boldsymbol\theta - \boldsymbol\theta_0}{|\boldsymbol\theta - \boldsymbol\theta_0|^2},
    \end{equation}
and rearrange to solve for the image position,
    \begin{equation}
    \boldsymbol\theta_{SMBH} \approx \boldsymbol\theta_0 - \tpm^2\frac{\boldsymbol\beta}{|\boldsymbol\beta|^2}.
    \end{equation}
The position of the SMBH image as a function of source impact
parameter is merely offset from the original location by the same
amount as the SMBH itself.  The magnification of the image,
    \begin{equation}
    \mu_{SMBH} \approx - \frac{|\boldsymbol\theta - \boldsymbol\theta_0|^4}{\tpm^4},
    \end{equation}
will also have the familiar form, but with respect to the new SMBH
location.

\subsection{Remarks}
Neither ellipticity nor external shear (nor, under proper
circumstances, an off-center SMBH) significantly change the
probability of observing lensing by a SMBH. The cross-sections,
optical depths, and statistics we calculated for the circularly
symmetric models, therefore, remain a reasonably good representation
even under these more general circumstances.  Other core images,
however, do not necessarily lie in the limiting case regions of the
image plane and may be affected by ellipticity and shear.  We cannot
assume that the results of the circularly symmetric calculations can
be applied to deduce their statistics for these configurations, but
postpone further analysis for future work.
%


%

\end{document}